\documentclass[fleqn,usenatbib]{mnras}

\usepackage{newtxtext,newtxmath}

\usepackage[T1]{fontenc}

\DeclareRobustCommand{\VAN}[3]{#2}
\let\VANthebibliography\thebibliography
\def\thebibliography{\DeclareRobustCommand{\VAN}[3]{##3}\VANthebibliography}

\usepackage{graphicx}	
\usepackage{amsmath}	
\usepackage{pdflscape}

\title[$T_e-T_e$ in diffuse gas]{Electron temperature relations in low metallicity, diffuse, and extraplanar gas of starburst galaxies}

\author[Hamel-Bravo et al.]{
Magdalena J. Hamel-Bravo,$^{1, 2}$
Deanne B. Fisher,$^{1, 2}$
Danielle A. Berg,$^{3}$
Alex J. Cameron,$^{4}$
John Chisholm,$^{3}$\newauthor
Glenn G. Kacprzak,$^{1,2}$
Barbara Mazzilli Ciraulo,$^{1,2}$
Anna F. McLeod,$^{5,6}$
Rodrigo Herrera-Camus$^{7}$
\\
$^{1}$Centre for Astrophysics and Supercomputing, Swinburne University of Technology, Hawthorn, Victoria 3122, Australia\\
$^{2}$ARC Centre of Excellence for All Sky Astrophysics in 3 Dimensions (ASTRO 3D), Australia\\
$^{3}$Department of Astronomy, University of Texas, Austin, TX 78712, USA\\
$^{4}$Sub-department of Astrophysics, University of Oxford, Keble Road, Oxford OX1 3RH, United Kingdom\\
$^{5}$Centre for Extragalactic Astronomy, Department of Physics, University of Durham, South Road, Durham DH1 3LE, UK\\
$^{6}$Institute for Computational Cosmology, Department of Physics, University of Durham, South Road, Durham DH1 3LE, UK\\
$^{7}$Departamento de Astronom\'ia, Universidad de Concepci\'on, Barrio Universitario, Concepci\'on, Chile\\
}

\date{Accepted XXX. Received YYY; in original form ZZZ}

\pubyear{\the\year{}}

\begin{document}
\label{firstpage}
\pagerange{\pageref{firstpage}--\pageref{lastpage}}
\maketitle

\begin{abstract}

In this work, we test the validity of $T_e$ - $T_e$ relations in resolved (10-200~pc) measurements of four nearby, low-metallicity (7.25 $\leq$ 12+log(O/H) $\leq$ 8.33), low-mass (10$^{6.78}$ $\leq$ M$_*$/M$_\odot$ $\leq$ 10$^{8.7}$), starburst (10$^{-4.5}$ $\leq$ sSFR $\leq$ 10$^{-0.3}$) galaxies. We obtain VLT/X-Shooter spectra of NGC~5253, NGC~0625, SBS~0335-052E and IC~2828, targeting regions within these galaxies with bright point-like sources and diffuse gas. Our observations are designed to extend from the galaxy midplane into extraplanar gas likely belonging to galactic winds. We measure electron temperatures from five different auroral lines: [\ion{N}{II}]~$\lambda$5755, [\ion{O}{II}]~$\lambda\lambda$7319,30, [\ion{S}{II}]~$\lambda\lambda$4069,76, [\ion{S}{III}]~$\lambda$6312, and [\ion{O}{III}]~$\lambda$4363. We compare the resulting $T_e$ - $T_e$ relations with previous studies of \ion{H}{II} regions in nearby spiral galaxies. Our results show that $T_e$ - $T_e$ relations in low-metallicity starburst galaxies do not significantly deviate from $T_e$ - $T_e$ relations in \ion{H}{II} regions of local spiral galaxies. We do not find significant differences in the diffuse, extraplanar gas. These results suggest that auroral lines provide a reliable metallicity diagnostic not only for high-redshift galaxies but also for the extended diffuse gas in extreme environments like outflows. 

\end{abstract}




\section{Introduction}

The chemical evolution of galaxies provides insight into the efficiency of star formation and the effects of feedback in the enrichment of the interstellar medium (ISM) and the circumgalactic medium (CGM) \citep{Roy1995, Tremonti2004, Tumlinson2017}. Furthermore, the ISM metallicity directly affects the properties of stellar populations, the impact of stellar feedback \citep{McLeod2021} and galaxy structures \citep{Gallazzi2005}. Observations of galaxy metallicities across cosmic time can help us understand how chemical elements build up in the Universe. From spatially resolved metallicity measurements of galaxies, we can understand the effects of different gas flows on the overall properties of galaxies. Thus, to understand the full picture of the chemical evolution of galaxies, we need a method that allows us to reliably measure metallicity across a broad range of environments \citep{Maiolino2019}. 

The ratio of auroral lines to nebular emission lines of the same ion is sensitive to the electron temperature ($T_e$) of the gas, which can be used to measure the metallicity. This ``direct'' method is frequently referred to as a ``gold standard'' in measuring gas-phase metallicity  \citep{Peimbert1967, Kewley2008, Berg2020}. This method, however, relies on high excitation emission lines that are 100 - 1000$\times$ fainter than more commonly measured strong lines, such as [\ion{O}{III}]~$\lambda$5007 and H$\alpha$. It has, therefore, been challenging to apply in low surface brightness environments, like diffuse extraplanar gas or in very high redshift galaxies. The recent development of high-throughput spectrographs on Keck and VLT, as well as the commissioning of \textit{JWST}, has not only led to an increase in studies of auroral lines, but also measurements beyond nearby \ion{H}{II} regions \citep{Cameron2021, Schaerer2022}. 

The use of auroral lines to calculate metallicity relies on the assumption that the ratio of a high-excitation line to a low-excitation line of a given ion accurately determines the $T_e$ of the gas, but there are some caveats and uncertainties that should be kept in mind when applying this method. The ionization structure of the gas can introduce uncertainties to the auroral line method. Gas with different ionization states can show a difference in $T_e$ \citep{Peimbert1967, Kewley2019}. To address these complexities, models such as the two-zone \citep{Lopez-Sanchez2012}, three-zone \citep{Garnett1992}, or four-zone ionization models \citep{Berg2021} are often used to separate the nebula into different ionization regions and measure $T_e$ for each. Ideally, to accurately measure the metallicity, one must determine the temperature structure of the gas by measuring $T_e$ from different ions that probe different ionization zones. The use of a single $T_e$ can bias the abundance measurement \citep[e.g.,][]{Arellano-Cordova2020, Rogers2022}.

Observations often rely on a single emission line ratio, such as [\ion{O}{III}]~$\lambda$4363/$\lambda$5007, to estimate electron temperatures. To account for unobserved ionization states when deriving metal abundances, empirical or photoionization model-based temperature relations ($T_e$ - $T_e$ relations) are commonly used. These relations are usually calibrated on observations of nearby \ion{H}{II} regions where multiple auroral lines are detectable and are generally well-behaved in such environments \citep{Esteban2009, Berg2020}. As a result, $T_e$ - $T_e$ relations are widely used when determining metallicities via the auroral line method, providing a practical approach for systems where only a single auroral line is available.

Metallicity measurements using auroral lines, particularly [\ion{O}{III}]~$\lambda$4363, and established $T_e$ - $T_e$ relations have been successfully applied to both resolved nearby systems \citep{Cameron2021,Hamel-Bravo2024} and integrated spectra of high-redshift galaxies \citep{Sanders2020, Laseter2024}. However, 
well-known differences --  including ionization state, electron density, and metallicity -- may lead to a change in the ionization structure \citep[e.g.][]{Berg2021}. This may lead to differences in the electron temperature relations between different ionization zones. Similar issues exist for extraplanar gas, in which large changes to the physical properties of the gas may likewise impact the ionization structure, and thus the applicability of the auroral lines to estimate metallicity. Ultimately, despite the frequent description of auroral lines as a gold standard tracer of metallicity, biases may still be present when applying this method in extreme environments. 


This work uses emission line spectra across the disks and extraplanar gas of four nearby, low-mass (< 10$^{8.7} M_{\odot}$), low-metallicity (12+log(O/H) < 8.35) galaxies to test the $T_e - T_e$ relationships in different environments. We detect multiple auroral lines to derive new electron-temperature relations for the diffuse and extraplanar gas. By examining $T_e$ - $T_e$ relationships in these high redshift analogues, in different regions of the galaxy, like the diffuse extraplanar gas and the bright sources in the disk, we can study how gas properties affect the direct method metallicities. 

The remainder of this paper is organized as follows: In Section~\ref{sec:data} we describe the data used in this work, including our galaxy sample (\S\ref{sec:galaxy sample}), our observations and data reduction (\S\ref{sec: Obs and reduction}), and the comparison sample (\S\ref{sec:comparison sample}). Section~\ref{sec: Emission lines} describes our method to measure emission line fluxes. In Section~\ref{sec:Te relations} we present our $T_e$ measurements, detailing our method used to derive them (\S\ref{sec:Te measurements}), the classification of point-like sources and diffuse gas (\S\ref{sec:point-sources_clasification}), the derived $T_e - T_e$ relations (\S\ref{sec:Te relations low} and ~\S\ref{sec:Te relations high}), a statistical comparison between the \ion{H}{II} regions sample and the diffuse gas (\S\ref{sec:HII_vs_diffuse}), and dependence on ionization, electron density and dust extinction (\S\ref{sec:Te relations trends}). Finally, in Section~\ref{sec: discussion}, we summarize our results and discuss their implications for metallicity measurements across different environments.

\section{Data}
\label{sec:data}
\begin{table}
\begin{tabular}{l c c c c}
\hline
 Galaxy & D     & log(M$_{\star}$) & log(SFR)     & 12 + log(O/H) \\ 
        & [Mpc] & [M$_\odot$]     & [M$_\odot$ yr$^{-1}$] & \\ 
 \hline
\multicolumn{1}{l}{NGC 5253} & 3.5 $^a$     & 8.64 $^c$       & -0.26 $^c$      & 8.19 $^f$ \\ 
\multicolumn{1}{l}{NGC 0625} & 4.0 $^a$       & 8.60 $^c$             & -1.20 $^c$    & 8.22 $^g$ \\ 
\multicolumn{1}{l}{SBS 0335-052E} & 59.0 $^b$     & 6.78 $^d$       & -0.15 $^e$     & 7.25 $^h$\\ 
\multicolumn{1}{l}{IC 2828}   & 13.0 $^b$    & 8.05 $^c$      & -1.45 $^c$     & 8.33 $^i$\\ \hline  
\label{tab:galaxy_prop}

\end{tabular}
\caption{Main properties of the galaxies. (a): Distances from the Cosmicflow database \protect\cite{Tully2016},(b): Distances from flow models \protect\cite{Kourkchi2020}, (c): \protect\citep{Marasco2023}, (d): \protect\cite{Reines2008},(e):\protect\citep{Herenz2023}, (f): \protect\cite{Skillman2003}, Metallicities measured using the direct method: (f):\protect\cite{Welch1970},(g):\protect\cite{Reines2008},(h):\protect\cite{Izotov2009},(i):\protect\cite{Brinchmann2008}}
\end{table}
\subsection{Galaxy sample}
\label{sec:galaxy sample}

We selected four galaxies for this study that are sufficiently nearby to resolve the diffuse gas from the bright sources, and that span a different range of properties than the spiral galaxies usually used in $T_e - T_e$ studies. The four nearby (3.5 Mpc $\leq$ D $\leq$ 59 Mpc), low-metallicity (7.25 $\leq$ 12+log(O/H) $\leq$ 8.4), low-mass (10$^{6.78}$ $\leq$ M$_*$/M$_\odot$ $\leq$ 10$^{8.7}$), starburst galaxies are NGC~5253, NGC~0625, SBS~0335-052E, and IC~2828. With our observations, taken under seeing conditions of $\sim$0.8\arcsec, we can obtain a spatial resolution of $\sim$20~pc for NGC~5353 \& NGC~0625, $\sim$50~pc IC~2828 and $\sim$250~pc for SBS0335-052E. Table~\ref{tab:galaxy_prop} summarizes the key properties of the sample. 

These galaxies have low masses, a clumpy structure, high star formation rates (-1.45 $\leq$ log(SFR) [M$_\odot$~yr$^{-1}$] $\leq$ -0.15), and low metallicities, similar to typical galaxies at higher redshift \citep{Izotov2021}. Therefore, we will use these results to inform how resolved electron temperature relationships, measured from different auroral lines, may behave in conditions similar to the early Universe. From H$\alpha$ imaging of these galaxies, we observe emission from ionized gas beyond the stellar disk, commonly referred to as diffuse extraplanar gas, indicating the presence of outflowing gas. To quantify the extent of our extraplanar measurements, we report the maximum projected distance of the X-Shooter slit from the disk along the minor axis in units of R$_{90}$, the radius enclosing 90\% of the galaxy's stellar light. For our sample, the maximum distances reached are: 0.8R$_{90}$ for NGC~5253, 0.7R$_{90}$ for NGC~0625, 4R$_{90}$ for SBS~0335-052 and 0.8R$_{90}$ for IC~2828. While most of our measurements lie just within the formal stellar extent, the lack of significant continuum emission and the vertical orientation of the structures support their interpretation as extraplanar gas.

NGC~5253 is a well-studied blue compact dwarf galaxy. The galaxy has a SFR that is 5$\times$ larger than the main-sequence value expected for a galaxy of this mass \citep{Popesso2023}. The metallicity is about $\sim 1/2$ the value for its stellar mass \citep{Calzetti2015,Tremonti2004}. The very young, massive clusters in the center of the galaxy produce a large ionizing photon flux ($Q(HI)\sim 7\times10^{52}$~s$^{-1}$), which is not well explained by the stellar mass of the central star-cluster ($\sim10^6$~M$_{\odot}$), and suggests a population of high-mass stars \citep{Smith2016}. Consistent with this, its spectra show evidence for Wolf-Rayet (WR) stars \citep{Schaerer1997}.  The harder ionization field could impact the ionization structure of the HII regions in this galaxy.  Moreover, it is similar to what is expected in high-redshift galaxies. \cite{Kobulnicky2008} shows in the H$\alpha$ and HI emission maps $\sim$0.8~kpc arc-like structures and filaments indicative of superbubbles and galactic winds.

SBS~0335-052E is a widely studied galaxy, as it is one of the most metal-poor starburst galaxies known \citep{Izotov1997}. These properties, along with its low mass (M$_{*}\sim10^7$~M$_{\odot}$), make it very useful as a nearby analog to the early universe. Spectra of SBS~0335-052E show that the galaxy is dominated by emission from high ionization ions in the optical range, such as [\ion{He}{II}]~$\lambda$4686 \citep{Kehrig2018}, [\ion{Ne}{V}]~$\lambda$3426 \citep{Thuan2005}, and [\ion{Ne}{V}]~$\lambda$14.32~$\mu$m \citep{Thuan2005,Mingozzi2025}. In particular, \cite{Kehrig2018} shows extended \ion{He}{II} up to a distance of $\sim$1.5~kpc from the brightest H$\alpha$ emission in the galaxy. Stellar population models struggle to model the strong radiation field required to observe these high ionization emission lines in SBS~0335-052 \citep{Wofford2021, Mingozzi2025}. \cite{Herenz2023} shows H$\alpha$ emission that extends to $\sim$15~kpc; the biconical shape of the emission and the connection to the starburst suggest this is an outflow.

NGC~0625 and IC~2828 are both low metallicity galaxies in which the star formation rate is more similar to the main-sequence value. As well, both have metallicity within 0.2~dex of the mass-metallicity relationship \citep{Tremonti2004}. IC~2828 shows plumes of ionised gas that are interpreted as the response of feedback processes \citep{Jaiswal2016}. NGC~625 does show spectral features of WR stars \citep{Monreal-Ibero2017}, and is described as having a galactic wind with a wind velocity of $\sim$200~km~s$^{-1}$ \citep{Marasco2023}. 

The overall sample of 4 targets are, therefore, selecting a range of star-forming activity in the low-mass, low-metallicity regime. All have evidence of young stellar populations that are impacting the diffuse ISM via feedback processes. Two of the targets are more extreme (SBS~0335-052 and NGC~5253), and two are more typical of low-mass galaxies (NGC~0625 and IC~2828). 


\subsection{Observation and data reduction}
\label{sec: Obs and reduction}

\begin{figure*}
    \centering
    \includegraphics[width=0.9\textwidth]{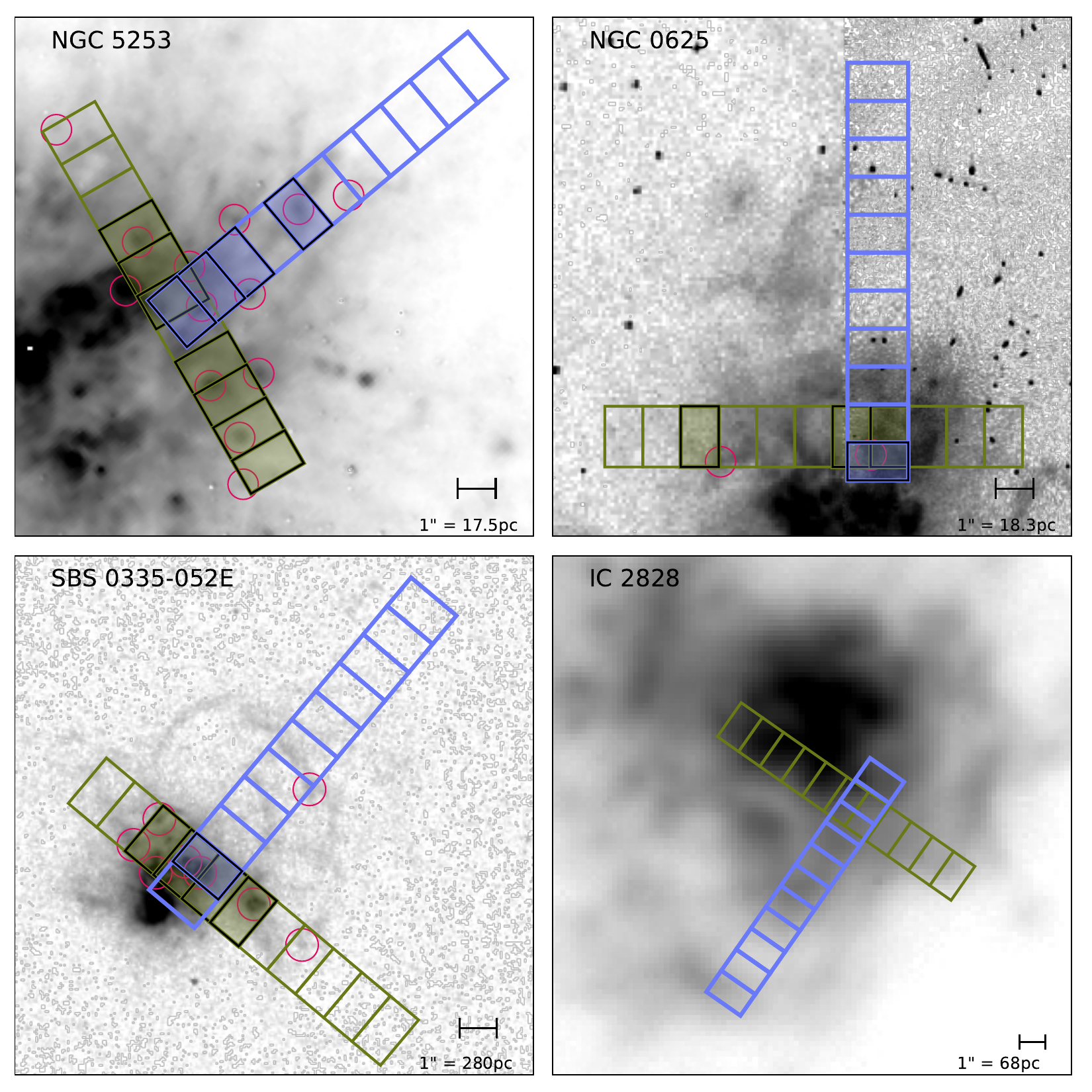} 
    \caption{Continuum subrtacted H$\alpha$ images of our four targets showing the position of the X-Shooter slits and the identified point-like sources of ionized gas. For NGC~5253 and NGC~0625 we use the continuum corrected {\it HST}/WFC2 F656N, for SBS~0335-052 we use {\it HST}/ACS F656N. For IC~2828 we continuum subtract the {\it VLT}/MUSE cube around the H$\alpha$ emission line, and then sum the remaining flux. Blue contours show the position of the minor axis slit and the 11 spatial elements within that slit. Green contours show the major axis slit. Red circles show the identified clumps with the adopted radius. Individual positions within slits that are colored correspond to positions with a contribution of clumps higher than 20\%. We only do this in the HST images given the clumps are unresolved in MUSE data.}
    \label{fig:slit_pos}
\end{figure*}

We observed these four galaxies using the X-Shooter spectrograph \citep{Vernet2011} on the ESO Very Large Telescope (VLT) between December 1st 2023 and April 5th 2024 as part of program ID~112.25V8.001, with a seeing between $\sim$0.6\arcsec\ and $\sim$1.1\arcsec. Each galaxy was observed with two different slit positions, one aiming at the galaxy disk and a second one aiming at a bright extended H$\alpha$ filament. Figure~\ref{fig:slit_pos} shows the positions of the two slits for each galaxy, overlaid on H$\alpha$ images. This figure will be discussed in more detail in Section~\ref{sec:point-sources_clasification}. We will refer to the slits as major axis slit (green slit in Figure~\ref{fig:slit_pos}) and minor axis slit (blue slit in Figure~\ref{fig:slit_pos}), respectively. The positions of the H$\alpha$ filaments were determined using archival MUSE data for each target. X-Shooter has three different arms, the UVB arm (3000 \AA - 5600 \AA), the VIS arm (5500 \AA - 10200 \AA) and the NIR arm (10200 \AA - 24800 \AA). All slits have a fixed length of 11\arcsec\ and we chose the broadest configuration for each arm (1.6\arcsec\ for the UVB, 1.5\arcsec\ for the VIS and 1.2\arcsec\ for the NIR). The average spectral resolution of the three arms is $\sim$~75~km/s. 

The observations were made in the mapping mode. We observed a pattern of object-sky-object. The exposure times varied for each target and wavelength arm (UVB, VIS, NIR). Individual sky frames were executed with equal exposure time to individual object frames. 

\begin{table}
\begin{tabular}{l c c c}
\hline
\multicolumn{4}{c}{Exposure times }                                                                        \\ \hline

& UVB   & VIS   & NIR                                                     \\
\hline
NGC~5253 minor axis      & 3 $\times$ 1560s  & 3 $\times$ 1500s & 3 $\times$ 1500s\\
                         & 1 $\times$ 1120s  & 1 $\times$ 920s  & 1 $\times$ 100s \\[.2cm]
NGC~5253 major axis      & 1 $\times$ 226s   & 1 $\times$ 250s  & 1 $\times$ 300s\\[.2cm]
NGC~0625 minor axis      & 3 $\times$ 1600s  & 3 $\times$ 1540s & 3 $\times$ 1500s\\
                         & 1 $\times$ 1108s  & 1 $\times$ 1162s & 1 $\times$ 960s \\[.2cm]
NGC~0625 major axis      & 1 $\times$ 226s   & 1 $\times$ 250s  & 1 $\times$ 300s\\[.2cm]
SBS~0335-052E minor axis & 4 $\times$ 1560s  & 4 $\times$ 1520s & 4 $\times$ 1450s\\
                         & 1 $\times$ 1140s  & 1 $\times$ 1148s & 1 $\times$ 1380s \\[.2cm]
SBS~0335-052E major axis & 1 $\times$ 226s   & 1 $\times$ 250s  & 1 $\times$ 300s\\[.2cm]
IC~2828 minor axis       & 4 $\times$ 1600s  & 4 $\times$ 1560s & 4 $\times$ 1600s\\
                         & 1 $\times$ 1160s  & 1 $\times$ 1158s & 1 $\times$ 1050s \\[.2cm]
IC~2828 major axis       & 1 $\times$ 226s   & 1 $\times$ 250s  & 1 $\times$ 300s\\[.2cm]

\hline  
\label{tab:exposure_times}
\end{tabular}
\caption{Summary of X-Shooter exposure times for each slit position and each arm (UVB, VIS and NIR). The format shows: (Number of OBs) $\times$ (Total exposure time per OB).} 
\end{table}
We performed one observing block (OB) for the major axis slit and 4 to 5 OBs for the minor axis slit for each galaxy. The exposure times and number of OBs per target were determined using the ESO exposure time calculator to be able to detect the faint auroral lines and not saturate bright emission lines. Exposure times for each slit are listed in Table~\ref{tab:exposure_times}. Despite our efforts to determine appropriate exposure times, some positions in the slits of NGC~5253 and NGC~0625 have saturated pixels at the position of the [\ion{O}{III}]~$\lambda$5007 and the [\ion{O}{III}]~$\lambda$4959 emission lines. This can be a problem when estimating the electron temperature from the [\ion{O}{III}]~$\lambda$4636 / $\lambda$5007 ratio, we address this in Section~\ref{sec: Emission lines}. 

The data reduction was done using the ESO Recipe Flexible Execution Workbench (Reflex) \citep{Freudling2013}. We reduce each OB individually and from each exposure, we subtract the closest in time sky exposure. The pipeline provides flux-calibrated and sky-subtracted 2D images of each exposure. Exposures were then combined using an inverse variance weighted average. 

We extracted individual spectra from the 2D reduced images by dividing the image in the spatial direction into 11 spatial elements of 1\arcsec\ each. This gave us 11 spectra along each slit. Figure~\ref{fig:slit_pos} displays the positions of the slits, each with 11 spatial elements overlaid on H$\alpha$ images of the galaxies.

\subsection{Comparison sample}
\label{sec:comparison sample}
We use \ion{H}{II} regions drawn from the CHemical Abundances Of Spirals (CHAOS) \citep{Berg2015} sample as a comparison. They measure $T_e$ from multiple auroral lines in a large number of \ion{H}{II} regions in nearby spiral galaxies using the Mutli-Object Double Spectrographs on the Large Binocular Telescope. We use data from NGC~0628 \citep{Berg2015} , NGC~5194 \citep{Croxall2015} , NGC~5457\citep{Croxall2016}, NGC~3184 \citep{Berg2020} and M~33 \citep{Rogers2022}. The CHAOS sample is significantly more massive (M$_{*}\sim 10^{10} - 10^{11}$~M$_{\odot}$) and significantly more metal-rich (12+log(O/H) $\sim 8.5-9.0$) than our sample. The comparison, therefore, allows us to determine if the $T_e-T_e$ relations of typical, local Universe spirals extend to higher $T_e$.

\section{Emission line measurements}
\label{sec: Emission lines}

The integrated spectra in each spatial element are corrected for Milky Way extinction by using the \cite{Cardelli1989} extinction law with $A_V$ values from the NASA/IPAC Extragalactic Database (NED, http://ned.ipac.caltech.edu) for each galaxy. We then estimate the stellar continuum. For spectra with continuum signal-to-noise ratio (SNR)~$>$~5 (calculated in the wavelength range from 4500~\AA\ to 4640~\AA\ where the flux is dominated by stellar continuum), we perform a stellar continuum fit using the Penalized PiXel-Fitting (pPXF) \citep{Cappellari2017}. We mask all visible emission lines and perform the fit using stellar models from E-Miles \citep{Vazdekis2016}, BPASS \citep{Eldridge2017} and GALAXEV \citep[][BC03]{Bruzual2003}. BPASS has the advantage of including binary evolution, while E-Miles and GALAXEV do not. However, from visual inspection, models from GALAXEV resulted in the lowest residuals around the H$\gamma$ emission line, so we adopted these models for all SNR~$>$~5 continuum subtractions. For spectra with continuum SNR~$\leq$~5 we use a 2 degree polynomial fit for the continuum, after masking emission lines. $\sim$30\% of our spectra have a continuum SNR~$>$~5. Most of these high SNR spectra correspond to spatial elements in the minor axis slits, which have higher exposure times than the major axis slits, that lie close to the stellar components of the galaxies.

\begin{figure*}
    \centering
    \includegraphics[width=1\textwidth]{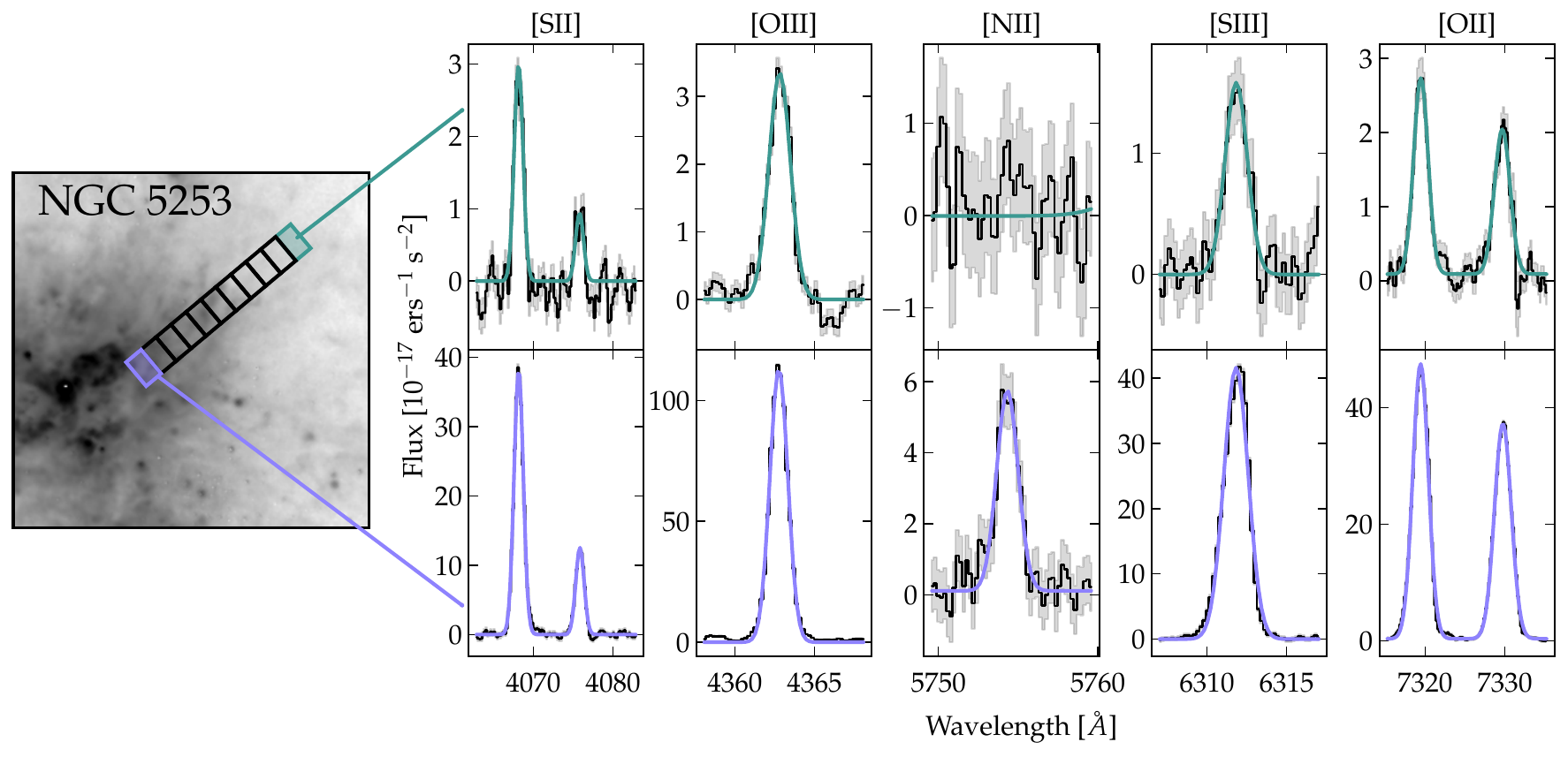} 
    \caption{Auroral emission lines for two different spatial elements in NGC~5253. The left panel shows an H$\alpha$ map of NGC~5253 with the X-Shooter spatial elements highlighted: purple, located closer to the H$\alpha$ flux peak and green, located further out in the diffuse extraplanar gas. Right panels show zoom-ins around the 5 auroral lines used in this work:  [\ion{S}{II}]~$\lambda\lambda$4068,76, [\ion{O}{III}]~$\lambda$4363, [\ion{N}{II}]~$\lambda$5755, [\ion{S}{III}]~$\lambda$6312, and [\ion{O}{II}]~$\lambda\lambda$7320,30. Top row shows emission line spectra for the spatial element in the extraplanar gas (green) and bottom panels correspond to the bright H$\alpha$ position (purple). In each panel, the black line shows the observed data, with the gray shaded region indicating the measured standard deviation in each wavelength channel. Gaussian fits to the emission lines are overplotted in green and purple.}
    \label{fig:spectra}
\end{figure*}

After continuum subtraction we fit relevant emission lines using a Gaussian model. For doublets we use a double Gaussian model with fixed wavelength offset between lines, this is the case for [\ion{O}{II}]~$\lambda\lambda$3726,29, [\ion{S}{II}]~$\lambda\lambda$4069,76 and [\ion{O}{II}]~$\lambda\lambda$7320,30. For H$\alpha$~$\lambda$6563, [\ion{N}{II}]~$\lambda$6548 and [\ion{N}{II}]~$\lambda$6584, we use a triple Gaussian model with fixed offset between centroids. The Gaussian fit includes a baseline offset, which accounts for imperfect continuum subtraction. We use the galaxy redshifts for an initial guess of the Gaussian centroid for each emission line and we set a minimum width of the Gaussian to match the instrument dispersion, $\sigma_{inst}~\sim$~0.3 \AA. We calculate the SNR in a wavelength range centered on the fitted Gaussian centroid and a width equal to the width of the Gaussian. We consider a line as detected if SNR~$\geq$~3. We estimate the flux uncertainty by performing 100 Monte Carlo simulations, perturbing the observed spectrum at each pixel based on its standard deviation, and taking the standard deviation of the resulting flux measurements. Figure~\ref{fig:spectra} shows an example of the fit of all five auroral lines for two different spatial elements in NGC~5253.

For some of our spatial elements, the [\ion{O}{III}]~$\lambda$5007 emission line is saturated. We inspect outputs from the reduction pipeline to determine which positions have saturated [\ion{O}{III}]~$\lambda$5007 emission. For those positions we use the theoretical ratio [\ion{O}{III}]~$\lambda$5007/$\lambda$4959~=~2.9 \citep{Osterbrock2006} to estimate the [\ion{O}{III}]~$\lambda$5007 flux. Some positions have saturated [\ion{O}{III}]~$\lambda$4959 emission as well, for those cases we can not make an estimate of the [\ion{O}{III}]~$\lambda$5007 flux, thus we can not calculate the electron temperature for [\ion{O}{III}]. 

Within the X-Shooter wavelength range, we have access to several hydrogen emission lines which we use to determine the dust extinction from Balmer line decrements. We fit all hydrogen lines from  H12~$\lambda$3250 to H$\alpha$~$\lambda$6563 and calculate their ratio relative to H$\beta$. We estimate the $A_V$ value by comparing the ratios with theoretical values assuming a $T_e = 10^4$~K and a $n_e = 100$~cm$^{-1}$ and the extinction curve for the Large Magellanic Cloud from \cite{Gordon2003}. We take the average $A_V$ value, weighted by the SNR of each line, for each position as the extinction value. Our estimated $A_V$ values range from 0.1 to 1, with a median $A_V$ of 0.3, which agrees with values from the literature for these galaxies. We correct the flux from each fitted emission line using the extinction curve for the Large Magellanic Cloud from \cite{Gordon2003}. For the rest of the paper, all properties are derived from the extinction-corrected emission line fluxes. Table~\ref{tab:emission_line} shows the reddening correction factors and fluxes for all measured lines for each slit position.

\begin{figure}
    \centering
    \includegraphics[width=\linewidth]{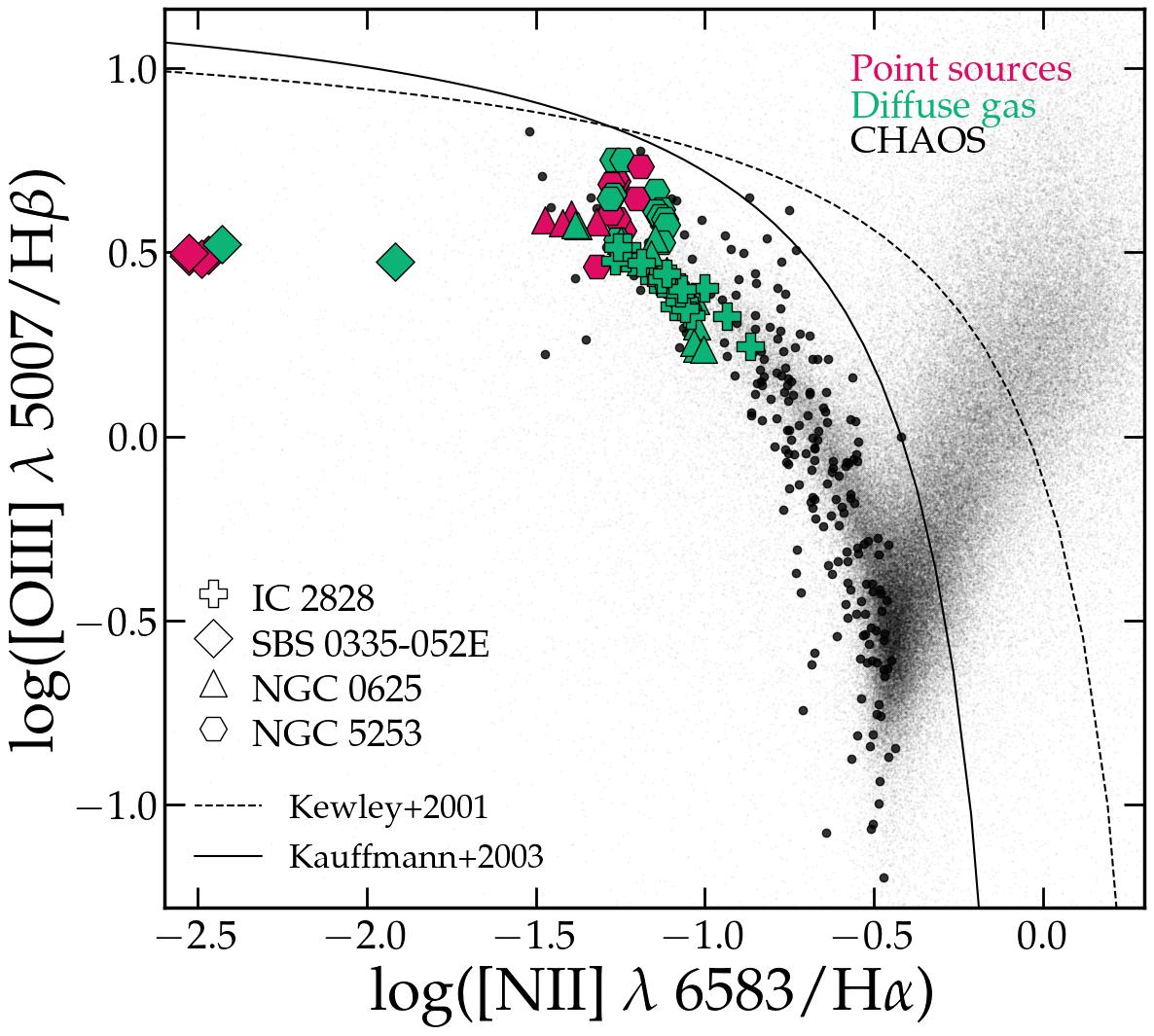}
    \caption{BPT diagram from the [\ion{O}{III}] $\lambda$5007 / H$\beta$ and the [\ion{N}{II}] $\lambda$6583 / H$\alpha$ ratio. Solid and dashed black lines show theoretical values from \protect\cite{Kewly2001} and \protect\cite{Kauffmann2003} respectively, below which emission line ratios are consistent with photoionization by stars. Pink symbols show measurements for our data that are classified as point-like sources and green symbols correspond to diffuse gas (see Section~\ref{sec:point-sources_clasification}). Each galaxy is shown with a different marker. Bigger solid black points show CHAOS data and smaller black dots show galaxies from the SDSS DR4 spectroscopic survey. }
    \label{fig:BPT_diagram}
\end{figure}

\section{Electron temperature relations}
\label{sec:Te relations} 

\subsection{Electron temperature measurements}
\label{sec:Te measurements}
We measure the electron temperature from 5 different auroral to nebular emission line ratios:

\begin{itemize}
  \item $T_e$ ([\ion{N}{II}]) : [\ion{N}{II}]~$\lambda$5755 / ($\lambda$6584 + $\lambda$6548)
  \item $T_e$ ([\ion{O}{II}]) : [\ion{O}{II}]~($\lambda$7319+$\lambda$7330) / ($\lambda$3726 + $\lambda$3729)
  \item $T_e$ ([\ion{S}{II}]) : [\ion{S}{II}]~($\lambda$4069+$\lambda$4076) / ($\lambda$6716 + $\lambda$6731)
  \item $T_e$ ([\ion{S}{III}]) : [\ion{S}{III}]~$\lambda$6312 / ($\lambda$9069 + $\lambda$9531)
  \item $T_e$ ([\ion{O}{III}]) : [\ion{O}{III}]~$\lambda$4363 / ($\lambda$5007 + $\lambda$4959).
\end{itemize}

We only considered emission lines with a SNR~>~3. For [\ion{S}{II}]~$\lambda\lambda$4069,76 and [\ion{O}{II}]~$\lambda\lambda$7319,30 if the fainter line in the doublet has a SNR < 3 but the brighter line has a SNR > 3 we use only the brighter line in our $T_e$ calculations. Both the [\ion{O}{II}]~$\lambda\lambda$3726,29 and the [\ion{S}{II}]~$\lambda\lambda$6716,31 doublets are sensitive to the electron density ($n_e$) of the gas for $n_e\sim10^1 - 10^4~$cm$^{-3}$ \citep{Osterbrock2006}. We derive the $n_e$ using both diagnostics, adopting a typical electron temperature of $T_e = 10^4$~K. However, we note that these diagnostics are largely insensitive to the assumed $T_e$. To estimate the uncertainty in $n_e$, we perform 1000 Monte Carlo realizations by perturbing the fluxes within their associated uncertainties. We then compute the variance in the resulting $n_e$ values and adopt this as the error on $n_e$. The resulting densities are reported in Table~\ref{tab:Te_values}.

We use the \texttt{getTemDen} function in \texttt{PyNeb} \citep{Luridiana2015} to calculate the $T_e$ from emission line ratios. We use the same atomic data as CHAOS (See Table~4 in \citep{Berg2015}) and the electron density calculated from the \ion{S}{II} doublet, for consistency with the CHAOS sample. For radiative transition probabilities of [\ion{O}{II}], [\ion{O}{III}] and [\ion{N}{II}] we use values from \cite{Fischer2004}, for [\ion{S}{II}] we use \cite{Mendoza2014} and for [\ion{S}{III}] we use \cite{Fischer2006}. For collision strength of [\ion{O}{II}] we use values from \cite{Kisielius2009}, for [\ion{O}{III}] we use \cite{Storey2014}, for [\ion{N}{II}] we use \cite{Tayal2011}, for [\ion{S}{II}] we use \cite{Tayal2010} and for [\ion{S}{III}] we use \cite{Hudson2012}. We estimate the uncertainties in $T_e$ using the same method as for $n_e$.

\subsection{Classification of diffuse regions and point-sources}
\label{sec:point-sources_clasification}

Historically, studies of $T_e$ either specifically target \ion{H}{II} regions \citep[e.g.][]{Berg2020} or alternatively entire galaxies. Our observations, however, find significant auroral line emission in both bright point-sources and regions that are more diffuse. As auroral lines are now being used to study gas that is clearly not in \ion{H}{II} regions, such as outflows and inflows \citep[e.g.][]{Cameron2021,Hamel-Bravo2024}, this study will be useful for testing for environmental differences. 

We use archival HST H$\alpha$+[NII] narrow-band images to identify brighter point-source regions that significantly impact the flux of our spectra. We expect these bright sources to have ionization structures similar to nearby \ion{H}{II} regions, thus we expect similar $T_e$ - $T_e$ relations to the bright \ion{H}{II} regions observed in CHAOS. For H$\alpha$ imaging, we use archival F656N filter data from {\it HST} for NGC~5253 (WFC3/UVIS P.I. Calzetti, program I.D. 6524), NGC~0625 (WFPC2 P.I. Skillman, program I.D. 8708) and SBS~0335-052E (ACS/WFC P.I. Oestlin, program I.D. 10575). There is no HST H$\alpha$ images available for IC~2828, and the MUSE observations do not use adaptive optics. We, therefore, cannot carry out the same identification in this target. For the analysis in the remainder of this section, we consider all spatial elements of IC~2828 as diffuse gas. This implies that our diffuse gas sample may have some contamination from spectra dominated by \ion{H}{II} regions. 

To estimate the stellar continuum in the F656N image, we use F547M, because this filter is primarily composed of starlight and is available for all sources. We calculate the distribution of the F656N/F547M ratio across the image. Regions with high ratios of F656N/F547M are more likely to be dominated by H$\alpha$ emission, while positions with low ratios are dominated by stellar continuum. We select regions that are representative of the continuum as those 5\% of pixels with the lowest F656N/F547M ratios. We then use those continuum-dominated pixels and fit a linear relation between F547M and F656N fluxes (F547M~=~a~$\times$~F656N~+~b). This relation provides the appropriate scaling factor between the two images. We then apply the scaling to each pixel in F547M to generate an F656N continuum image. Finally, we subtract the continuum image from the F656N image and obtain a continuum-subtracted H$\alpha$ image.

Figure~\ref{fig:slit_pos} shows the continuum-subtracted H$\alpha$ images. For IC~2828 we show an image created from archival {\it VLT}/MUSE data (P.I. Fossati, program I.D. 098.A-0364). To align X-Shooter slits and HST images we use acquisition images taken in the V band and align them with the HST F547M.

We identify point-sources in the continuum-subtracted H$\alpha$ images by visually locating peaks in flux. For each peak, we extract the H$\alpha$ flux profile as a function of distance from the brightest point and compute the mean flux in radial bins. We fit a Gaussian function to this profile and extract its maximum, width ($\sigma$), and baseline. We define the source radius as the Gaussian full width at half maximum (FWHM). We identify the scatter in the baseline flux around each point-source as the standard deviation at a distance of 2.5 - 3.5~$\times$ the $\sigma$ that represents the width of the Gaussian fit to the point-source. We then only include those sources in which the peak brightness is 3 times higher than the standard deviation of the background. This ensures that the flux profile decreases in all directions like a Gaussian up to a reasonable radius, helping to exclude flux peaks associated with extended ionized filaments.

We then calculate, for each spatial element in our slits, the ratio between the total H$\alpha$ flux and the flux that is inside 1~FWHM of the fitted Gaussian to the point-source. If the fraction is at least 20\%, we classify the spatial element as being significantly impacted by point-source emission. Figure~\ref{fig:slit_pos} shows the positions of these identified point-sources in our X-Shooter slits. Slit positions classified as affected by point-source emission are shaded in either blue (minor axis) and green (major axis).

In Figure~\ref{fig:BPT_diagram} we show the position of all our spatial elements in the BPT diagram, colored in pink if they are classified as positions with point-sources or green if they are classified as diffuse gas. For comparison, we also include the \ion{H}{II} regions from CHAOS in black. All of our measurements fall within the star-forming locus and are located toward the lower [\ion{N}{II}]/H$\alpha$ and higher [\ion{O}{III}]/H$\beta$ end of the CHAOS distribution.

\subsection{Low and intermediate ionization zones}
\label{sec:Te relations low}

\begin{figure*}
    \centering
    \includegraphics[width=0.8\textwidth]{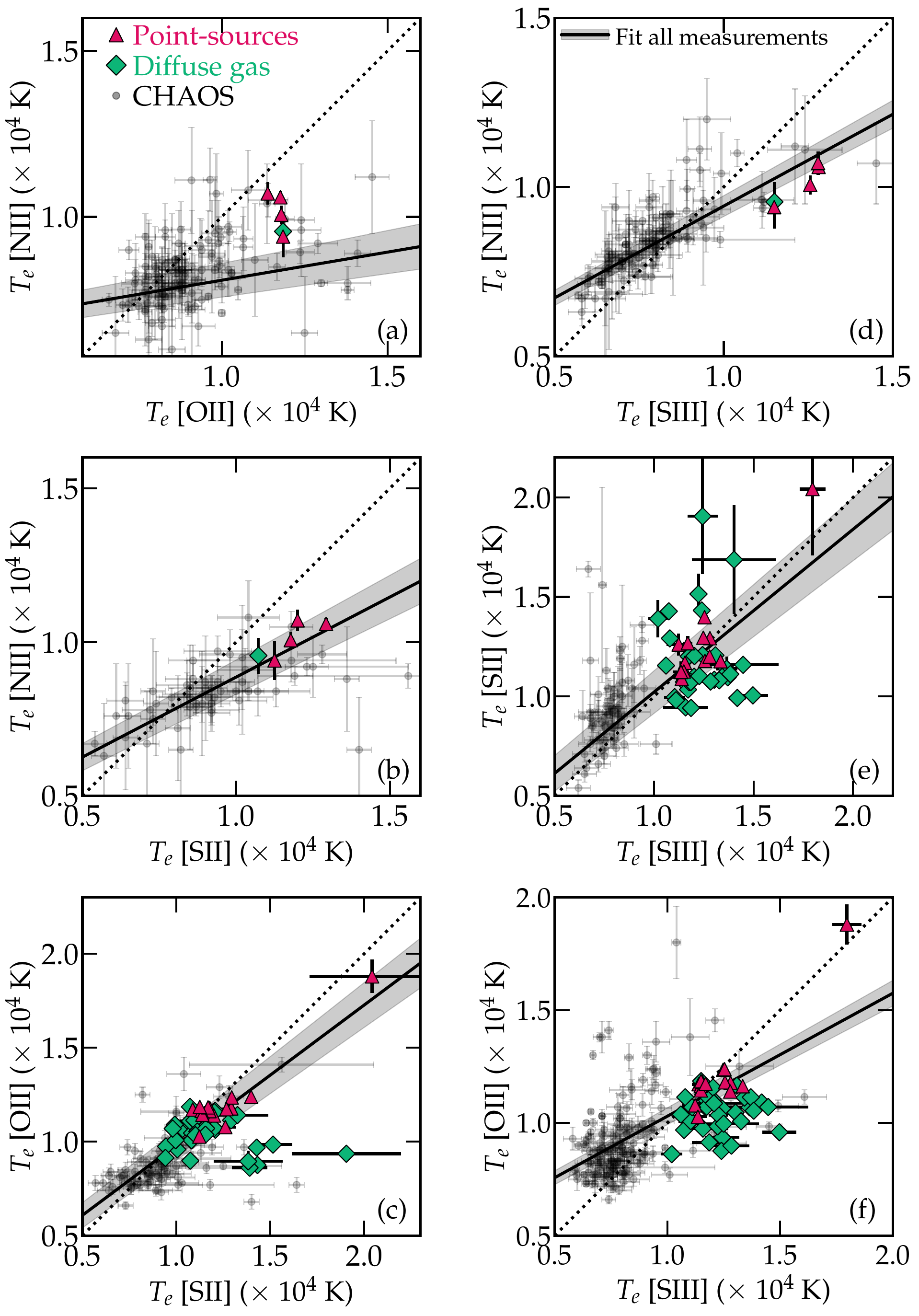}
    \caption{$T_e$ - $T_e$ relations for ions probing the low and intermediate ionization zone. Left column shows $T_e$ - $T_e$ relations between the low ionization ions ([\ion{N}{II}], [\ion{O}{II}] and [\ion{S}{II}]). The right column shows $T_e$ relations between the low ionization ions and [\ion{S}{III}]. Black dots show CHAOS data, pink triangles show bright point-like sources in our data and green diamonds show diffuse gas in our data. The black line shows the linear fit to all the data. }
    \label{fig:te-te_low_ion}
\end{figure*}

Assuming a three zone ionization structure of the gas, the electron temperature from the low ionization zone can be measured from [\ion{N}{II}], [\ion{S}{II}] or [\ion{O}{II}], the intermediate zone can be measured from [\ion{S}{III}] and the high ionization zone from [\ion{O}{III}] \citep{Berg2015}. With our X-Shooter data, we measure all emission lines necessary to estimate $T_e$ from these five ions. This allows us to study the relationship between $T_e$ diagnostics for the low ionization zone and also between $T_e$ diagnostics that sample different ionization zones. Using the classification we made in Section~\ref{sec:point-sources_clasification}, we can compare $T_e - T_e$ relations of \ion{H}{II} regions in CHAOS with our measurements of point-sources and diffuse gas. In this section, and Section~\ref{sec:Te relations high}, we examine where our data fall relative to CHAOS and other studies.  A more quantitative comparison is provided in Section~\ref{sec:HII_vs_diffuse}.

Figure~\ref{fig:te-te_low_ion} shows all $T_e - T_e$ relations for the low ionization zone and the intermediate ionization zone. The CHAOS data is shown in black dots, our positions with bright point-like sources are shown as pink triangles and our positions dominated by diffuse gas are shown as green diamonds. We fit linear relations, using an orthogonal distance regression routine from \texttt{SciPy} \citep{Virtanen2020}, to all the data. The linear fit is plotted as a solid black line, with a shaded region indicating the uncertainty of the fit. Table~\ref{tab:te-te_fits} shows the linear equations of the fits that include all data, corresponding to the black line in $T_e$ - $T_e$ plots. We also calculate the root mean square (RMS) for the CHAOS data, the point-like sources and the diffuse gas around the fitted line and report these values in Table~\ref{tab:te-te_fits}. For $T_e - T_e$ relations including [\ion{N}{II}] we do not have enough measurements in our point-like sources or diffuse gas to estimate an RMS, so we do not repost these values.  


For [\ion{N}{II}], [\ion{S}{II}], and [\ion{O}{II}], we expect the $T_e$ – $T_e$ relations to follow a 1:1 trend due to their similar ionization potentials \citep{Garnett1992}. On the left column of Figure~\ref{fig:te-te_low_ion} we show $T_e$ - $T_e$ relations between these three low ionization zone tracers. We have five  [\ion{N}{II}]~$\lambda$5755 detections in our data, one for a position of diffuse gas and four for point-like sources. Panels (a) and (b) in Figure~\ref{fig:te-te_low_ion} show the $T_e$([\ion{O}{II}]) - $T_e$([\ion{N}{II}]) and the $T_e$([\ion{S}{II}]) - $T_e$([\ion{N}{II}]) relationships respectively. For our sample $T_e$([\ion{N}{II}]) is on average 1700~K lower than both $T_e$([\ion{O}{II}]) and $T_e$([\ion{S}{II}]). Even though we expect them to show a 1:1 trend, $T_e$([\ion{N}{II}]) tends to show lower values than $T_e$([\ion{O}{II}]) and $T_e$([\ion{S}{II}]) in previous studies \citep{Esteban2009, Pilyugin2009, Berg2020}.  This discrepancy between $T_e$ values has been attributed to density inhomogeneities in the gas \citep{Mendez-Delgado2023, RickardsVaught2023}. The [\ion{O}{II}] and [\ion{S}{II}] nebular emission lines have lower ($\sim$ 10$^3$~cm$^{-3}$) critical densities than the [\ion{N}{II}] nebular emission line ($\sim$ 8~$\times$~10$^4$~cm$^{-3}$), while all three auroral lines have a high critical density ($>$~10$^6$~cm$^{-3}$). So, in the case of density inhomogeneities, if a fraction of the emitting gas has a $n_e > 10^3$~cm$^{-3}$, the nebular emission lines of [\ion{S}{II}] and [\ion{O}{II}] may be suppressed due to collisional de-excitation. This can result in artificially high $T_e$ estimates. In Section~\ref{sec:Te relations trends} we investigate possible effects of $n_e$ in $T_e - T_e$ relations.

Panel (c) in Figure~\ref{fig:te-te_low_ion} shows the $T_e$([\ion{S}{II}]) - $T_e$([\ion{O}{II}]) relation. We have a higher number of $T_e$ measurements for this relationship, 15 for point-like sources, 33 for diffuse gas and 79 for CHAOS. Most of our measurements are within the uncertainties of the linear fit, except for five measurements that clearly show an excess of $T_e$([\ion{S}{II}]) over $T_e$([\ion{O}{II}]). These five outliers are from diffuse gas in NGC~5253 and NGC~0625, which are the two galaxies with the highest spatial resolution ($\sim$20~pc). Factors that could affect these $T_e$ measurements are the ionization of the gas, the electron density, or the extinction correction \citep{RickardsVaught2023}. In Section~\ref{sec:Te relations trends} we explore these possibilities. We include one measurement from SBS~0335-052E in the $T_e$([\ion{S}{II}]) - $T_e$([\ion{O}{II}]) relation. This data point, located at $T_e \sim 20000 K$ - approximately 10$^4$~K higher than the rest of the sample- remains consistent with the linear fit despite its significantly elevated electron temperature. The RMS scatter around the fitted relation is $\sim$600~K for the point-like source measurements, and $\sim$1700~K for the diffuse gas measurements. The latter is comparable to the RMS of $\sim$1500~K observed in the CHAOS dataset.

On the right panels of Figure~\ref{fig:te-te_low_ion} we show the $T_e$ - $T_e$ relations for the three ions probing the low ionization zone with [\ion{S}{III}], which is probing the intermediate ionization zone. The difference in ionization potential between these ions could lead to a difference in $T_e$. Panel (d) in Figure~\ref{fig:te-te_low_ion} shows the $T_e$([\ion{S}{III}]) - $T_e$([\ion{N}{II}]) relation. A tight correlation between $T_e$[\ion{N}{II}] and $T_e$[\ion{S}{III}] has been found previously \citep{Berg2020, RickardsVaught2023}. Due to our low number of $T_e$([\ion{N}{II}]) measurements we can not robustly confirm this. Our five measurements, however, are consistent with the CHAOS data. 

Panel (e) shows the $T_e$([\ion{S}{III}]) - $T_e$([\ion{S}{II}]) relation. We have 15 measurements for point-like sources, 30 for diffuse gas and 65 for CHAOS. Our linear fit is consistent with previous work \citep{RickardsVaught2023}. The RMS of the diffuse gas sample is double that of the point-source sample, as reported in Table~\ref{tab:te-te_fits}. This trend shows the largest RMS for $T_e$ relations between low and intermediate ionization zone. We have one measurement for SBS~0335-052E with $T_e$([\ion{S}{III}])$ = 1.8 \times 10^4$~K and $T_e$([\ion{S}{II}])$ = 2.0 \times 10^4$~K that is consistent with the linear fit within measurement uncertainties. 

Panel (f) in Figure~\ref{fig:te-te_low_ion} shows the $T_e$([\ion{S}{III}]) - $T_e$([\ion{O}{II}]) relation. We have 15 measurements for point-like sources, 30 for diffuse gas and 166 for CHAOS.  While the CHAOS data show systematically higher $T_e$([\ion{O}{II}]) than $T_e$([\ion{S}{III}]), our measurements exhibit the opposite trend, with $T_e$([\ion{S}{III}]) exceeding $T_e$([\ion{O}{II}]). Our linear fit shows a more rapid increase of $T_e$([\ion{S}{III}]) over $T_e$([\ion{O}{II}]). Using spectra of \ion{H}{II} regions in nearby galaxies with stellar masses $10^{9.4} \leq M_{*} \leq 10^{10.8}$, \cite{RickardsVaught2023} found the opposite trend for these two $T_e$ indicators, with some \ion{H}{II} regions showing similar results to ours. We have one measurement for SBS~0335-052 at $T_e \sim 19000$ K, which is not consistent with our linear fit, but it is consistent with the 1:1 line. This relation shows the lower scatter of $T_e$ relations between low ionization and [\ion{S}{III}]. Previous studies have identified [\ion{N}{II}] as the preferred tracer of low-ionization gas due to the tight correlation it shows with tracers of higher ionization gas \citep{Berg2020}. In low metallicity galaxies, [\ion{N}{II}] is less abundant, thus less likely to be detected. Given that [\ion{O}{II}] shows the lowest scatter with [\ion{S}{III}], $T_e$([\ion{O}{II}]) would be a better tracer in these systems. [\ion{O}{II}] has shown a large scatter in previous studies \citep{Berg2020}, but we are finding good trends for our low-metallicity systems.

\subsection{High ionization zone}
\label{sec:Te relations high}
\begin{figure*}
    \centering
    \includegraphics[width=0.8\textwidth]{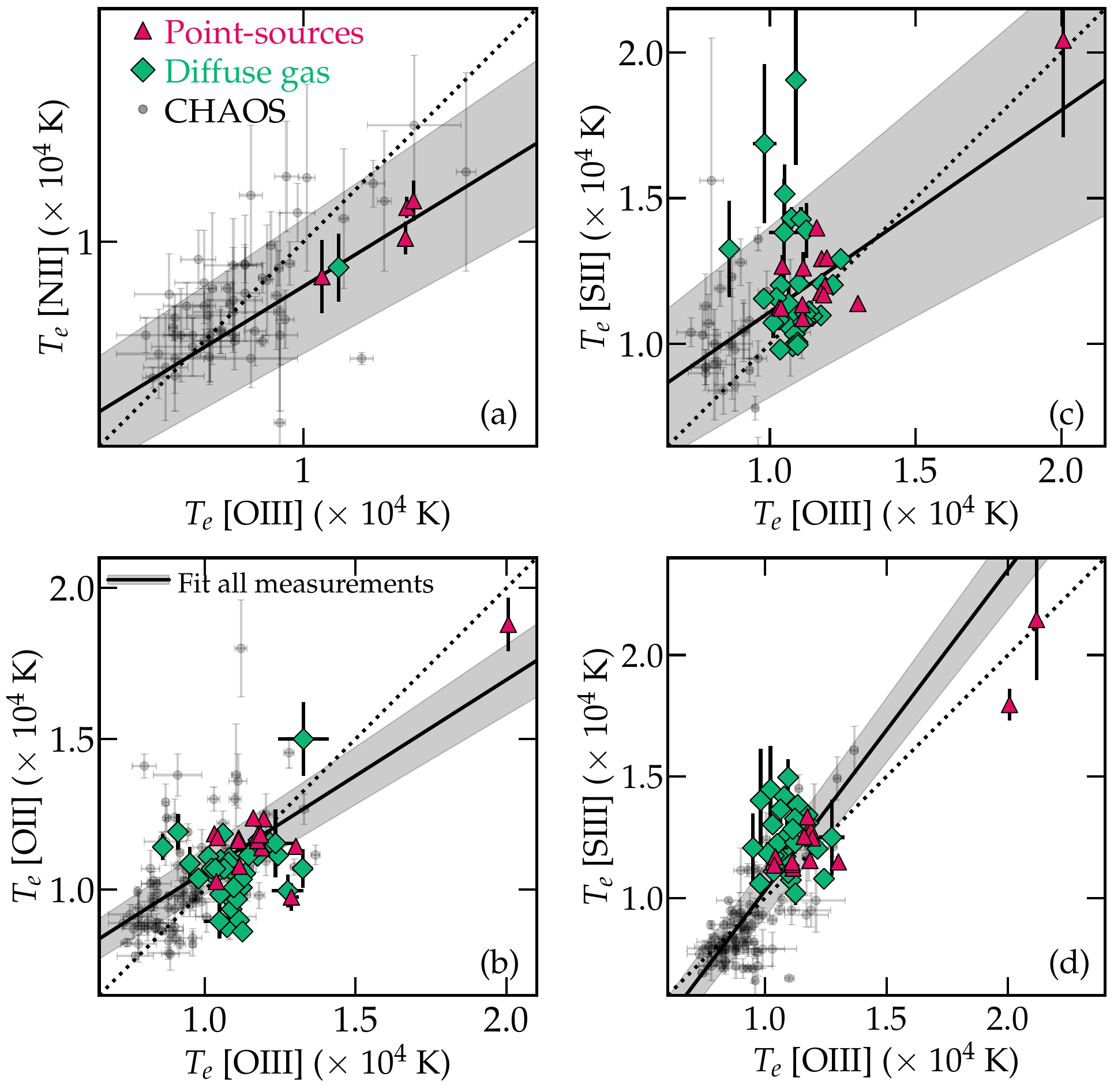}
    \caption{$T_e$ - $T_e$ relations for [\ion{O}{III}] with all lower ionization zone ions ([\ion{N}{II}], [\ion{O}{II}], [\ion{S}{II}] and [\ion{S}{III}]). Black dots show CHAOS data, pink triangles show bright point-like sources in our data and green diamonds show diffuse gas in our data. The black line shows the linear fit to all the data. }
    \label{fig:te-te_high_ion}
\end{figure*}

The electron temperature calculated from [\ion{O}{III}] traces the high ionization zone. In Figure~\ref{fig:te-te_high_ion} we compare $T_e$([\ion{O}{III}]) with all the other $T_e$. The [\ion{O}{III}]~$\lambda$4363 auroral line is the most used when measuring the metallicity of both nearby and high-redshift galaxies. This is because it is the brightest and easiest to observe with its corresponding strong line at $\lambda$5007. For this reason, the $T_e$ - $T_e$ relations for this ion are of particular interest to understand the possible systematics associated with using the direct method. 

The $T_e$([\ion{O}{III}]) - $T_e$([\ion{N}{II}]) relation is shown in panel (a) of Figure~\ref{fig:te-te_high_ion}. Our measurements are in the higher $T_e$ range of the CHAOS measurements, and show the same trend of slightly lower $T_e$([\ion{N}{II}]) compared with $T_e$([\ion{O}{III}]). We show the $T_e$([\ion{O}{III}]) - $T_e$([\ion{O}{II}]) relation in panel (b). Our measurements are very consistent with the measurements from CHAOS. For SBS~0335-053E, we measure $T_e$([\ion{O}{II}])$~=~18794$K and $T_e$([\ion{O}{III}])$~=~20062$~K, which is consistent with our linear fit. Panel (c) shows the $T_e$([\ion{O}{III}]) - $T_e$([\ion{S}{II}]) relation. We observe a large scatter for this relation, with some points showing an excess $T_e$([\ion{S}{II}]) over $T_e$([\ion{O}{III}]), which could be due to the uncertainties associated with measuring $T_e$([\ion{S}{II}]) mentioned previously. For SBS~0335-052E we measure consistent values of $T_e$([\ion{O}{III}]) and $T_e$([\ion{S}{II}]). 

Previous studies have shown a large discrepancy between [\ion{O}{III}] and [\ion{S}{III}] \citep{Perez-Montero2006, Berg2015}, which is attributed to observational uncertainties. The consensus is that [\ion{O}{III}] is less reliable than [\ion{S}{III}] in the higher ionization zone due to the large scatter found in $T_e$ relations involving $T_e$([\ion{O}{III}]). In panel (d) of Figure~\ref{fig:te-te_high_ion} we show the $T_e$ - $T_e$ relation for [\ion{O}{III}] and [\ion{S}{III}]. The RMS of our measurements seems very consistent with CHAOS. We have two measurements for SBS~0335-052E, one with $T_e$([\ion{O}{III}])$~=~20062$~K and $T_e$([\ion{S}{III}])$~=~17962$~K, and the second one with $T_e$([\ion{O}{III}])$~=~21173$~K and $T_e$([\ion{S}{III}])$~=~21463$~K. The second one is consistent with the 1:1 line and considering error bars it is also consistent with our linear fit. The first one is not consistent with our linear fit.  

Overall, we observe a higher scatter in the diffuse gas than in the point-like sources in all the $T_e - T_e$ relations. This could be because our emission line measurements in the diffuse gas have a lower SNR than in the bright point-like sources, which can result in noisier measurements. There could also be a systematic offset of diffuse gas measurements compared to \ion{H}{II} regions. We explore systematic offsets between the different samples in the next section.

{\renewcommand{\arraystretch}{1.5}
\begin{table*}
\begin{center}
\begin{tabular}{ccccccc}
\hline
\multicolumn{7}{c}{$T_e$ - $T_e$ relations linear fits}                                                                        \\ \hline
\multicolumn{3}{c|}{y = m x + b}           & \multicolumn{1}{c|}{} & \multicolumn{3}{c}{RMS}                         \\ \cline{1-3} \cline{5-7} 
x - y     & m   & \multicolumn{1}{c|}{b}   & \multicolumn{1}{c|}{} & Point-sources & Diffuse gas & CHAOS \ion{H}{II} regions \\ \cline{1-3} \cline{5-7} 

$T_e$([\ion{O}{II}]) - $T_e$([\ion{N}{II}])        & 0.17 $\pm$ 0.04 & \multicolumn{1}{c|}{0.64 $\pm$ 0.04} & \multicolumn{1}{c|}{} & --           & --         & 0.1                  \\ 
$T_e$([\ion{S}{II}]) - $T_e$([\ion{N}{II}])        & 0.48 $\pm$ 0.04 & \multicolumn{1}{c|}{0.4 $\pm$ 0.04} & \multicolumn{1}{c|}{} & --           & --         & 0.12                  \\ 
$T_e$([\ion{S}{II}]) - $T_e$([\ion{O}{II}])        & 0.73 $\pm$ 0.05 & \multicolumn{1}{c|}{0.26 $\pm$ 0.07} & \multicolumn{1}{c|}{} & 0.06           & 0.17         & 0.15                  \\ 
$T_e$([\ion{S}{III}]) - $T_e$([\ion{N}{II}])        & 0.54 $\pm$ 0.02 & \multicolumn{1}{c|}{0.4 $\pm$ 0.02} & \multicolumn{1}{c|}{} & --           & --         & 0.06                  \\ 
$T_e$([\ion{S}{III}]) - $T_e$([\ion{S}{II}])        & 0.88 $\pm$ 0.08 & \multicolumn{1}{c|}{0.11 $\pm$ 0.1} & \multicolumn{1}{c|}{} & 0.09           & 0.18         & 0.18                  \\ 
$T_e$([\ion{S}{III}]) - $T_e$([\ion{O}{II}])        & 0.55 $\pm$ 0.03 & \multicolumn{1}{c|}{0.47 $\pm$ 0.03} & \multicolumn{1}{c|}{} & 0.11           & 0.13         & 0.14                  \\ 
$T_e$([\ion{O}{III}]) - $T_e$([\ion{N}{II}])        & 0.61 $\pm$ 0.08 & \multicolumn{1}{c|}{0.31 $\pm$ 0.08} & \multicolumn{1}{c|}{} & --           & --         & 0.07                  \\ 
$T_e$([\ion{O}{III}]) - $T_e$([\ion{O}{II}])        & 0.64 $\pm$ 0.06 & \multicolumn{1}{c|}{0.4 $\pm$ 0.06} & \multicolumn{1}{c|}{} & 0.08           & 0.11         & 0.13                  \\ 
$T_e$([\ion{O}{III}]) - $T_e$([\ion{S}{II}])        & 0.91 $\pm$ 0.15 & \multicolumn{1}{c|}{0.15 $\pm$ 0.17} & \multicolumn{1}{c|}{} & 0.07           & 0.18         & 0.16                  \\ 
$T_e$([\ion{O}{III}]) - $T_e$([\ion{S}{III}])        & 1.28 $\pm$ 0.07 & \multicolumn{1}{c|}{-0.24 $\pm$ 0.08} & \multicolumn{1}{c|}{} & 0.11           & 0.1         & 0.09                  \\ 
\hline
\end{tabular}
\caption{$T_e$ - $T_e$ linear fit results. These are the fit to all the data, thus the black line in Figures~\ref{fig:te-te_low_ion}~\&~\ref{fig:te-te_high_ion}. The first column shows the ions used in the $T_e$ - $T_e$ relation. The second column shows the slope and the third column shows the intercept. The last three columns show the RMS around the fitted line of the three different groups of measurements: bright point-sources, diffuse gas and CHAOS, respectively. \label{tab:te-te_fits}}
\end{center}
\end{table*}}

\subsection{$T_e - T_e$ in \ion{H}{II} regions vs. diffuse gas}
\label{sec:HII_vs_diffuse}

The aim of this work is to compare our new measurements with previous measurements from bright \ion{H}{II} regions in CHAOS. This comparison is challenging because the two datasets sample different ranges of $T_e$ and differ in size. The CHAOS sample has more than double the number of measurements compared to our point-sources and diffuse gas samples. Fitting linear relations to each of the three samples and comparing them is further complicated by the small range of $T_e$ values covered by our sample compared to CHAOS (see Figures~\ref{fig:te-te_low_ion}~\&~\ref{fig:te-te_high_ion}). A straightforward hypothesis is that the emission line flux from the X-Shooter spectrum in the regions with point-sources is dominated by the \ion{H}{II} regions, and thus follow similar $T_e - T_e$ relations as CHAOS. We therefore fit a single linear relation to the combined data. We bin the data in uniform steps of 2000~K in the x-coordinate for each relationship. The detailed procedure for binning and fitting is described in Appendix~\ref{append:binned te-te}. Figures~\ref{fig:te-te_low_ion_bin}~\&~\ref{fig:te-te_high_ion_bin} show the same as Figures~\ref{fig:te-te_low_ion}~\&~\ref{fig:te-te_high_ion} respectively, with the bins and the fit to the bins. 

To examine if the diffuse gas sample deviates significantly from this relation, we calculate the mean residual ($\langle r \rangle$) of the sample. We do this by calculating the orthogonal distance of our measurements to the fit and then calculating the mean for each of our samples (CHAOS, point-sources and diffuse gas). We compare $\langle r \rangle$ in the CHAOS sample and point-source sample to that of the diffuse gas sample. A large $\langle r \rangle$ value, compared to the average measurement error of $T_e$, would imply that the diffuse gas sample significantly deviates from the \ion{H}{II} region $T_e - T_e$ relations. We compare the mean residual to the mean uncertainty of the measured $T_e$ ($\sigma_{Te}$) in each sample. We also calculate the RMS of the residuals for each sample. This allows us to explore how much scatter is in each group.

We report the measured $\langle r \rangle$, RMS and $\sigma_{Te}$, for each $T_e - Te$ relation in Figures~\ref{fig:te-te_low_ion_bin}~\&~\ref{fig:te-te_high_ion_bin}. For most $T_e - T_e$ relations the mean residual of the diffuse gas sample is less or equal than three times the mean $T_e$ uncertainty ($\leq$~3$\sigma_{Te}$). This indicates that the diffuse gas sample does not show a statistically significant deviation from the $T_e - T_e$ relations defined by \ion{H}{II} regions. The only exception is the $T_e([\ion{O}{III}]) - T_e([\ion{S}{III}])$ relation, shown in panel (d) of Figure~\ref{fig:te-te_high_ion_bin}, where the mean residual is 4.3 times larger than the mean measurement uncertainty. In the next section we explore physical properties of the gas that could be impacting this $T_e - T_e$ relation.

For all $T_e - T_e$ relations, and for all three samples, the RMS of the residuals is larger than the average measurement uncertainty. This suggests that the scatter cannot be fully explained by measurement errors alone. Possible explanations include intrinsic scatter in the $T_e - T_e$ relations, underestimated uncertainties, or limitations in the adopted linear model. In the next section, we examine some physical properties that could contribute to the scatter.

\subsection{Electron temperature relations trends}
\label{sec:Te relations trends}
\begin{figure*}
    \centering
    \includegraphics[width=1\textwidth]{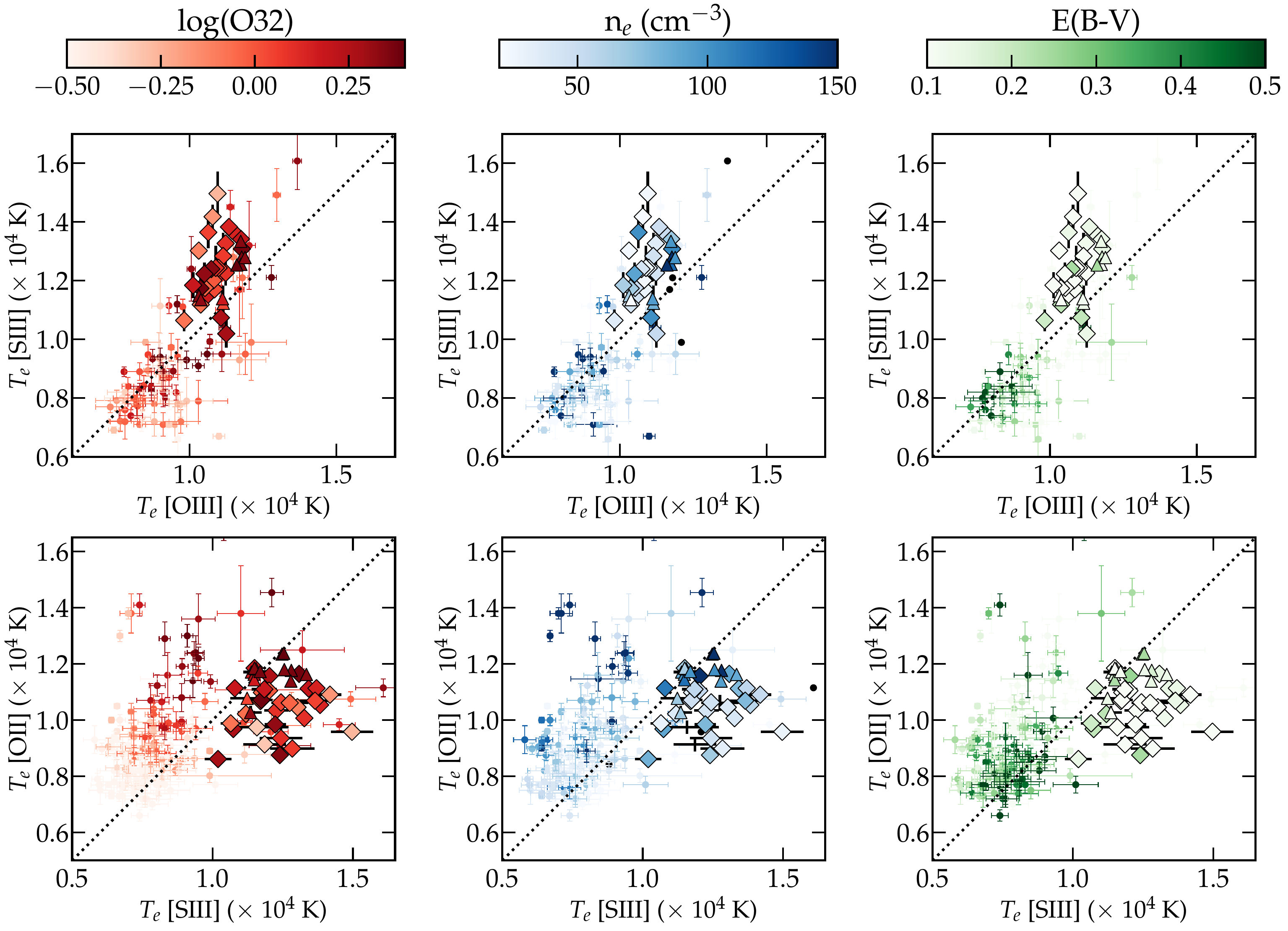}
    \caption{$T_e$ - $T_e$ relations colored by O32, $n_e$ and E(B-V). Top panels show the $T_e$([\ion{O}{III}]) - $T_e$([\ion{S}{III}]) relation, bottom panels show $T_e$([\ion{S}{III}]) - $T_e$([\ion{O}{II}]). Left panels are colored by O32 in red, middle panels are colored by $n_e$ in blue and right panels are colored by E(B-V) in green. Small circles show CHAOS data, triangles show our measurements for point-like sources, and diamonds show our measurements for diffuse gas.}  
    \label{fig:te-te_O3_ne_EBV}
\end{figure*}

In Figure~\ref{fig:te-te_O3_ne_EBV} we show two examples of $T_e$ - $T_e$ relations and color them by three different parameters to explore if some of these properties could be driving the scatter in these $T_e$ - $T_e$ relations. In the first row we show $T_e$([\ion{O}{III}]) - $T_e$([\ion{S}{III}]) (high ionization zone -intermediate ionization zone) and in the bottom row we show $T_e$([\ion{S}{III}]) - $T_e$([\ion{O}{II}]) (intermediate ionization zone - low ionization zone). In the left column, we color these relations by O32 = [\ion{O}{III}]~$\lambda$5007/[\ion{O}{II}]~$\lambda\lambda$3726,29  in red, in the middle column we color them by electron density in blue, and in the right column by color excess (E(B-V)) in green. We exclude SBS~0335-052E from these plots because it has a much higher $T_e$ than the rest of the sample, to better show the color gradients.

\subsubsection{Ionization}

The left column of Figure~\ref{fig:te-te_O3_ne_EBV} shows the $T_e$ - $T_e$ relations colored by O32, which is a proxy for ionization of the gas. The multi-ionization zone model assumes well-defined distinct ionization structures within the gas. This structure can vary with the ionization of the gas, leading to different $T_e$ - $T_e$ relations that depend on the ionization.

In the left columns of Figure~\ref{fig:te-te_O3_ne_EBV}, \ion{H}{II} regions with higher ionization are shown to have higher $T_e$, consistent with the expectations from multi-ionization zone models. Several studies have previously investigated the dependency of $T_e$ - $T_e$ relations with O32 \citep[e.g.][]{Berg2020, Yates2020, RickardsVaught2023}. They found a dependence of $T_e$ with ionization, where \ion{H}{II} regions with higher ionization tend to have higher $T_e$. This trend is expected, as regions with higher ionization are typically exposed to harder radiation fields, which deposit more energy into the gas and result in higher $T_e$.

It is not clear from Figure~\ref{fig:te-te_O3_ne_EBV} that there is a difference in the dependence of $T_e$([\ion{O}{III}]) - $T_e$([\ion{S}{III}]) (high ionization zone - intermediate ionization zone) and $T_e$([\ion{S}{III}]) - $T_e$([\ion{N}{II}]) (intermediate ionization zone - low ionization zone) with ionization in our data. 

\cite{Berg2020} found less scatter around the correlation in the $T_e$([\ion{S}{III}]) - $T_e$([\ion{N}{II}]) relation for \ion{H}{II} regions with low ionization, while \ion{H}{II} regions with high ionization had lower scatter in the $T_e$([\ion{O}{III}]) - $T_e$([\ion{S}{III}]).

From the top left panel in Figure~\ref{fig:te-te_O3_ne_EBV} we observe that most measurements that are close to the 1:1 line correspond to higher ionization, while measurements that are the furthest from the 1:1 line correspond to lower ionization. \cite{RickardsVaught2023} and \cite{Yates2020} also found that the highest deviation from the 1:1 line between temperatures of the low and the high ionization zones occurs in \ion{H}{II} regions with low O32 ratios. They observed an excess of $T_e$([\ion{O}{III}]) over $T_e$([\ion{S}{III}]) for \ion{H}{II} regions with $O32 < 0.5$. Most of our sample has an $O32 > 0.5$ and we observe the opposite trend, $T_e$([\ion{S}{III}]) exceeds $T_e$([\ion{O}{III}]).

\subsubsection{Electron density}

The effect of $n_e$ on $T_e - T_e$ relations has also been studied \citep[e.g.][]{Mendez-Delgado2023, RickardsVaught2023}, since it is expected to cause differences in electron temperatures from different ions. This is because collisional de-excitation affects some emission lines more than others, making some emission lines appear fainter, which can bias emission line ratios and thus electron temperature measurements. In the middle column panels of Figure~\ref{fig:te-te_O3_ne_EBV}, we plot $T_e$([\ion{O}{III}]) - $T_e$([\ion{S}{III}]) (top panel) and $T_e$([\ion{S}{III}]) - $T_e$([\ion{O}{II}]) (bottom panel) colored by $n_e$ derived from the [\ion{S}{II}]~$\lambda\lambda$6716,31 doublet, in blue. 

Due to the lower critical density of [\ion{O}{II}]~$\lambda\lambda$3726,29 (10$^3$~-~10$^4$~cm$^{-3}$) compared to the auroral [\ion{O}{II}]~$\lambda\lambda$7320,30 ($\sim$10$^6$ cm$^{-3}$) line, the $T_e$([\ion{O}{II}]) is particularly sensitive to $n_e$. In particular, in higher $n_e$ environments, we could overestimate $T_e$([\ion{O}{II}]) if we do not apply $n_e$ corrections or underestimate $n_e$. We include $n_e$ measurements in our $T_e$ calculations, and we measure a low $n_e$ ($n_e < 250$~cm$^{-3}$) for all our spatial elements, but we can not discard electron density inhomogeneities at lower spatial scales. The middle column bottom panel of Figure~\ref{fig:te-te_O3_ne_EBV} shows the $T_e$([\ion{S}{III}]) - $T_e$([\ion{O}{II}]) relation colored by $n_e$. We observe a trend of higher $n_e$ values having a higher $T_e$[\ion{O}{II}]. This could be consistent with an underestimation of $n_e$ in these higher $n_e$ environments. \cite{RickardsVaught2023} found that the [\ion{O}{II}] and [\ion{S}{II}] density diagnostics could be biased towards lower $n_e$ and this could explain why we observe higher $T_e$ from [\ion{O}{II}] compared to other ions in higher density environments. 

\subsubsection{Dust extinction}

Since $T_e$ measurements rely on emission line ratios, they can be affected by reddening corrections. The larger the wavelength difference between the auroral line and the nebular emission line is, the more significant this effect becomes. For the five auroral lines measured in this work, [\ion{O}{II}] should be the one most affected by the extinction correction. An over estimation of E(B-V) would lead to an under estimation of the [\ion{O}{II}]~($\lambda$7319+$\lambda$7330) / ($\lambda$3726 + $\lambda$3729) ratio which would lead to an under estimation of $T_e$([\ion{O}{II}]). In this work we estimate E(B-V) from multiple hydrogen line ratios to H$\beta$ (see Section~\ref{sec: Emission lines}), which is the same procedure as the one used in CHAOS. 

In the right column of Figure~\ref{fig:te-te_O3_ne_EBV} we plot $T_e$ relations for our data plus the CHAOS data colored by E(B-V) in green. We observe a relation between E(B-V) and $T_e$ in both the $T_e$([\ion{S}{III}]) - $T_e$([\ion{O}{II}]) and the $T_e$([\ion{O}{III}]) - $T_e$([\ion{S}{III}]), where measurements with higher extinction tend to have lower $T_e$ values. This is expected because higher metallicity galaxies (lower $T_e$) tend to have more dust \citep{Remy-Ruyer2014}. 

We observe that positions with higher E(B-V) values tend to lie closer to the 1:1 line. In the bottom panel, we observe that for our data we measure a lower $T_e$ from [\ion{O}{II}] relative to [\ion{S}{III}] for measurements with low E(B-V) values. This could mean that we are overestimating the reddening for these positions, however, this is unlikely due to the low estimated values of $A_V$ (See Section~\ref{sec: Emission lines}).

\section{Summary and discussion}
\label{sec: discussion}

Using the VLT/X-Shooter spectrograph we measure the electron temperature from 5 different auroral lines ([\ion{N}{II}]~$\lambda$5755, [\ion{O}{II}]~$\lambda\lambda$7319,7330, [\ion{S}{II}]~$\lambda\lambda$4069,4076,[\ion{S}{III}]~$\lambda$6312,[\ion{O}{III}]~$\lambda$4363) for different positions within low metallicity starburst galaxies (sample shown in Fig.~\ref{fig:slit_pos}). As illustrated in the figure, our analysis includes regions that are point-like sources, presumably dominated by \ion{H}{II} regions, as well as regions dominated by diffuse gas, both in the ISM and extraplanar. As we show in Fig.~\ref{fig:te-te_low_ion}~\&~\ref{fig:te-te_high_ion}, we compare $T_e$ - $T_e$ relations in these different environments (point-sources versus diffuse gas) within the galaxies of our sample. Our resolved, low-metallicity measurements are likewise compared to $T_e$ - $T_e$ relations from CHAOS \citep{Berg2015,Berg2020}. CHAOS observations only target \ion{H}{II} regions in nearby spiral galaxies, thus more comparable to our point-source classified measurements. 

As expected for low metallicity environments, our galaxies have higher median $T_e$ values than the CHAOS galaxies. Using the [\ion{O}{III}] auroral line we find that the median $T_e$ in CHAOS is $0.93\times10^4$~K, compared to $1.3\times10^4$~K for our sample. Despite these differences, we find that $T_e$ - $T_e$ relations in our metal-poor, high-ionization regions do not significantly deviate from CHAOS.


We use the CHAOS + low-metallicity sample to search for differences in the behavior of $T_e-T_e$ relationships based on properties of the ISM. In Fig.~\ref{fig:te-te_O3_ne_EBV} we show examples of $T_e - T_e$ relations compared to variation in O32, electron density and extinction. Lower ionization (traced by lower O32) systems tend to have lower $T_e$ values. This is true for both the low ionization zone ([\ion{O}{II}]) and the high ionization zone ([\ion{O}{III}]). We do not find that this relation increases scatter in the $T_e$ - $T_e$ relations. This behavior has been observed in previous studies on $T_e$ relations \citep{Berg2020}, our result extends this to lower metallicity. 

Electron density variations are proposed to contribute to scatter in $T_e$ relations. This is due to the fact that increases of $n_e$ do not impact all ions equally. For example, higher electron density will drive fainter [\ion{O}{II}]~$\lambda\lambda$~3736,39 emission, which is then measured as a higher $T_e$. This effect is not as pronounced for [\ion{O}{III}] and [\ion{S}{III}]. The combined difference leads to scatter in the $T_e$ - $T_e$ relations for these species \citep{Mendez-Delgado2023,RickardsVaught2023}. We do, indeed, observe such a trend in our sample where positions with lower $n_e$ show lower $T_e$([\ion{O}{II}]) in comparison to $T_e$([\ion{S}{III}]) than those positions with higher $n_e$. 

There is a very strong trend of the lowest-metallicity galaxies having much less dust \citep[e.g.][]{Fisher2014}, and thus less extinction. We clearly see this trend in Fig.~\ref{fig:te-te_O3_ne_EBV}, where our sample shows lower values of color excess relative to CHAOS. We do not see any significant offset of $T_e - T_e$ relations for different E(B-V) probed in CHAOS or our sample. We note that this includes [\ion{O}{II}], which is heavily dependent on extinction law modeling. 

Overall, our results show that $T_e$ correlations are robust across a wide variety of environments. This implies that use of these optical auroral lines as metallicity tracers is likewise robust. We discuss the implications for some of those applications. 

\subsection{Metallicity measurements at high redshift}

High-redshift metallicities are critical to study the early evolution of galaxies. Using JWST the [\ion{O}{III}]~$\lambda$4363 auroral line is more easily observed at redshifts up to $z\sim10$ \citep{Curti2023, Katz2023,Laseter2024, Morishita2024}. In most cases, only [\ion{O}{III}] is observed. These observations then must relate [\ion{O}{III}]~$\lambda$4363 to the lower ionization zones, and therefore require $T_e$ - $T_e$ relations established at low redshift to derive a total O/H abundance. However, the stellar populations and ISM of the typical high-redshift galaxy differ significantly from that of the typical low-redshift galaxy. At higher redshift, galaxies tend to be clumpier \citep{Guo2015, Carniani2018}, have lower metallicities, have higher electron densities and are more highly ionized \citep{Paalvast2018, Cameron2023, Sanders2023}.
A common practice is to specifically target lower redshift galaxies with analogous ISM conditions as tests for those conditions in the higher redshift universe \citep[e.g.][]{Izotov2021}. Our results are thus informative to current work at $z \gtrsim 4$. 

Within our sample of galaxies, SBS~0335-052E shows the most extreme conditions. It is the most metal poor, 12+log(O/H) $\approx$ 7.3 \citep{Papaderos2006}, very low mass, (M$_{star} \sim 10^7$~M$_{\odot}$), and highest ionization in our sample  \citep[O32$\approx$17;][]{Izotov1997}. These values resemble properties of galaxies found at $z \sim 5 - 7$. \cite{Cameron2023} found that more than half of their sample of $5.5 < z < 8.5$ galaxies observed with JWST have O32$ > 10$. Similarly, \cite{Tang2023} found a sample of galaxies at $z \sim 7 - 9$ to have 12+log(O/H) $\sim 7.84$ and having extremely high ionization, with O32 values of $\sim$ 17.84. We can therefore use SBS~0335-052 as a characteristic galaxy to search for the kinds of deviations expected in the early Universe, as others have in previous work\citep[e.g.][]{Kehrig2018,Herenz2023}. We find that this target follows the same trends in $T_e - T_e$ as the rest of our sample, as well as the CHAOS data. We find very little difference between the bright point-sources in SBS~0335-052 from the diffuse gas, down to $\sim$ 280~pc resolution. 

Especially useful are the relations that include [\ion{O}{III}] electron temperatures, shown in Figure~\ref{fig:te-te_high_ion}. This is because such relations allow the estimation of the $T_e$ in the low-ionization zone from $T_e$([\ion{O}{III}]), typically the brightest auroral line, in galaxies with extreme properties. Our results suggest that we can rely on commonly used $T_e$ - $T_e$ relations for metallicity measurements. Recent work, at more modest redshift of $z \sim 2 - 3$, finds similarly that $T_e-T_e$ relationships for full galaxies are not different from those of the local Universe \citep{Cataldi2025}. It, therefore, follows that our works supports the use of metallicity calculations based on the ionization corrections of local Universe galaxies for those at $z > 4$.

\subsection{Metallicity measurements in the diffuse gas}

Electron temperature relations are sensitive to the ionization structure of the gas, since different ions trace different ionization zones with potentially different electron temperatures. Despite the complex structure of \ion{H}{II} regions, they are usually well described by a sphere with shells of different ionization zones \citep{Stromgren1939, Morisset2016, Berg2021}. This structure forms because photons with high energy are absorbed closer to the ionizing source and lower energy photons reach the outer layers. Outside \ion{H}{II} regions, the ionization structure may differ significantly. In the diffuse ISM, or extraplanar gas, the ionization structure is much less constrained and thus may be very different from \ion{H}{II} regions. Different physical processes such as radiation leakage \citep{Chisholm2018}, gas flows \citep{Veilleux2005} and/or turbulent mixing \citep{Scalo2004} can disrupt the ISM leading to spatial variations in ionization and temperature. These variations could result in $T_e - T_e$ relations being different outside of \ion{H}{II} regions, and thus resulting in inaccurate metallicity measurements when using a single auroral line.

Using HST H$\alpha$ imaging, we identify regions dominated by bright point-like sources and regions of diffuse gas (Section~\ref{sec:point-sources_clasification}). Our results show that diffuse gas measurements do not significantly deviate from $T_e - T_e$ relations established by \ion{H}{II} regions. This is shown by the comparable residuals for the diffuse gas and \ion{H}{II} regions, to the $T_e - T_e$ relations established by \ion{H}{II} regions (Figures~\ref{fig:te-te_low_ion_bin}~\&~\ref{fig:te-te_high_ion_bin}). Also, from $T_e - T_e$ relations established by the whole sample, we observe that the diffuse gas has a higher scatter around fitted relations than the position with bright point-sources. That scatter, however, is comparable to the scatter of the CHAOS data, which only includes \ion{H}{II} regions (See Table~\ref{tab:te-te_fits}). This implies that $T_e - T_e$ relations remain valid outside of \ion{H}{II} regions, and furthermore that metallicity measurements from a single auroral line, like [\ion{O}{III}], in regions with no bright sources are likewise valid.  

\cite{Cameron2021} and \cite{Hamel-Bravo2024} used the [\ion{O}{III}]~$\lambda$4363 auroral line to measure the metallicity of the outflows of low metallicity starburst galaxies, Mrk~1486 and NGC~1569, respectively. These studies show the first spatially resolved observations of metal-loading in outflows, which is a fundamental parameter for galaxy evolution \citep[e.g][]{Sharda2021}. Similar to high redshift galaxies, measuring multiple auroral lines in the diffuse gas is challenging, thus we need $T_e$ - $T_e$ relations to account for unobserved ionization zone. The lack of significant difference in our scaling relationships suggests that single ion methods to derive the metallicity from $T_e$ is stable in outflows.  

Each spatial element in our work is 1\arcsec\ $\times$ 1.5{\arcsec}. Due to the different distances of the galaxies the extracted spectra are sampling different physical sizes, from $\sim$17~pc in NGC~5253 to $\sim$230~pc in SBS~0335-052. We do not observe any trends between the physical sizes we are probing and the electron temperature relations. This implies that $T_e$ - $T_e$ relations hold at small scales of $\sim$17~pc and also when integrating over a bigger surface of $\sim$230~pc. This is an indication that $T_e$ - $T_e$ relations yield accurate results even when observing the integrated spectra of high redshift galaxies. Small-scale variations, at least at the scales probed in this work, do not seem to affect the overall temperature structure of the gas. 

Overall, our work supports the use of single ion $T_e$ measurements to characterize the properties of the ISM in a range of environments, beyond the typical \ion{H}{II} region in local spirals. This includes in both the high density extreme regions of high-z galaxies, and the diffuse gas expelled from outflows.

\section*{Acknowledgements}

Parts of this research were supported by the Australian Research Council Centre of Excellence for All Sky Astrophysics in 3 Dimensions (ASTRO 3D), through project number CE170100013.  Based on observations collected at the European Organisation for Astronomical Research in the Southern Hemisphere under ESO programme(s) 112.25V8.001. Based on data obtained from the ESO Science Archive Facility.

\section*{Data Availability}

All raw data files are accessible in the ESO Science Archive\footnote{https://archive.eso.org/eso/eso\_archive\_main.html}.  The data underlying this article will be shared on reasonable request to the PI, Magdalena Hamel Bravo at mhamelbravo@swin.edu.au.



\bibliographystyle{mnras}
\bibliography{example} 




\appendix

\section{Emission line measurements and physical conditions}

In Table~\ref{tab:emission_line}~and~\ref{tab:Te_values} we report the measured properties from the spatial elements in our data. Table~\ref{tab:emission_line} shows the measured emission line fluxes ratios to H$\beta$ normalized such that  H$\beta$=100. Table~\ref{tab:Te_values} shows the electron temperature and electron density measurements.

\begin{landscape}
\renewcommand{\tabcolsep}{6pt}
\begin{table}
\begin{tabular}{ccccccccccccc}
\hline
\multicolumn{12}{c}{Observed line ratios [F(H$\beta$) = 100]}                                                                        \\ \hline
Galaxy & Position & E(B$-$V) & F [\ion{O}{II}]  & e [\ion{O}{II}] & F [\ion{O}{II}]  & e [\ion{O}{II}]  &  F H12th  & e H12th  & F H11th   & e H11th  & F H10th  & e H10th \\
       &          &     & $\lambda$3726    & $\lambda$3726 & $\lambda$3729 & $\lambda$3729 & $\lambda$3750 & $\lambda$3750 & $\lambda$3770 & $\lambda$3770 & $\lambda$3798 & $\lambda$3798 \\
 \hline
NGC 5253 & Min - 1 & 0.176& 74.57 & 0.059 & 95.729 & 0.067 & 2.569 & 0.015 & 3.246 & 0.014 & 4.425 & 0.015\\
NGC 5253 & Min - 2 & 0.125& 81.369 & 1.666 & 110.168 & 2.908 & 2.721 & 0.013 & 3.551 & 0.017 & 4.753 & 0.017\\
NGC 5253 & Min - 3 & 0.139& 84.446 & 0.303 & 115.165 & 0.509 & 2.656 & 0.017 & 3.472 & 0.02 & 4.723 & 0.025\\
NGC 5253 & Min - 4 & 0.141& 85.099 & 1.77 & 116.589 & 2.995 & 2.63 & 0.021 & 3.425 & 0.021 & 4.635 & 0.022\\
NGC 5253 & Min - 5 & 0.131& 93.994 & 1.043 & 128.828 & 2.362 & 2.685 & 0.027 & 3.467 & 0.029 & 4.689 & 0.039\\
NGC 5253 & Min - 6 & 0.136& 105.934 & 0.187 & 146.645 & 0.234 & 2.573 & 0.051 & 3.34 & 0.063 & 4.819 & 0.077\\
\hline

\end{tabular}
\caption{Observed, non extinction corrected, emission line ratios normalized such that F(H$\beta$) = 100. The first column lists the galaxy name, and the second column indicates the position within the galaxy, labeled as `Maj-N` or `Min-N` for extractions along the major and minor axis, respectively. The third column gives the E(B$-$V) reddening value derived from the Balmer decrement. Each emission line measurement is reported in two columns: the flux ratio relative to H$\beta$, and the corresponding uncertainty. All values are unitless and represent relative intensities. (This table is available in its entirety in machine-readable form.)}
\phantomsection
\label{tab:emission_line}
\end{table}

\begin{table}
\begin{tabular}{lcccccccccccccccc}
\hline
Galaxy & Position & Type & $n_e$ & e $n_e$ & $n_e$ & e $n_e$ & $T_e$ & e $T_e$ & $T_e$ & e $T_e$ & $T_e$ & e $T_e$ & $T_e$ & e $T_e$ & $T_e$ & e $T_e$ \\

 &  &  & [\ion{O}{II}] & [\ion{O}{II}] & [\ion{S}{II}] & [\ion{S}{II}] & [\ion{N}{II}] & [\ion{N}{II}] & [\ion{O}{II}] & [\ion{O}{II}] & [\ion{S}{II}] & [\ion{S}{II}] & [\ion{S}{III}] & [\ion{S}{III}] & [\ion{O}{III}] & [\ion{O}{III}] \\
\hline
NGC 5253 & Min-1 & Point-like source & 162.0 & 1.0 & 191.0 & 2.0 & 10591 & 184 & 11771 & 15 & 12926 & 31 & 12806 & 22 & 11776 & 5 \\
NGC 5253 & Min-2 & Point-like source & 101.0 & 20.0 & 121.0 & 2.0 & 10064 & 282 & 11798 & 78 & 11778 & 41 & 12556 & 36 & 11750 & 6 \\
NGC 5253 & Min-3 & Point-like source & 94.0 & 3.0 & 99.0 & 2.0 & 10707 & 345 & 11390 & 26 & 12000 & 52 & 12790 & 46 & 11889 & 7 \\
NGC 5253 & Min-4 & Point-like source & 89.0 & 18.0 & 107.0 & 3.0 & -- & -- & 11645 & 70 & 12064 & 50 & 13082 & 44 & 11782 & 7 \\
NGC 5253 & Min-5 & Diffuse gas & 87.0 & 12.0 & 98.0 & 3.0 & -- & -- & 11625 & 61 & 11766 & 69 & 13342 & 66 & 11744 & 10\\
NGC 5253 & Min-6 & Point-like source & 78.0 & 1.0 & 70.0 & 4.0 & -- & -- & 11123 & 53 & 10979 & 93 & 13422 & 131 & 11750 & 17 \\
\hline
\end{tabular}
\caption{Physical conditions of the ionized gas for each extracted region. The first column lists the galaxy name, followed by the position along the slit (column 2), labeled as `Maj-N` or `Min-N` for spatial elements along the major or minor axis, respectively. Column 3 indicates the morphological classification of the region, as either `Point-like` or `Diffuse`. Columns 4–7 report the electron densities (in cm$^{-3}$) derived from the [\ion{O}{II}] and [\ion{S}{II}] doublets with their uncertainties. Columns 8-17 show electron temperatures (in K) derived from the [\ion{N}{II}], [\ion{O}{II}], [\ion{S}{II}], [\ion{S}{III}], and [\ion{O}{III}] diagnostic lines, respectively. (This table is available in its entirety in machine-readable form.)}
\phantomsection
\label{tab:Te_values}
\end{table}
\end{landscape}


\section{Binned $T_e$ - $T_e$ relations}

In order to estimate how much our $T_e$ measurements in diffuse gas deviate from $T_e$ relations established from \ion{H}{II} regions, we first establish the $T_e - T_e$ relations in \ion{H}{II} regions. We make the assumption that our point-like source measurements are dominated by \ion{H}{II} regions, and therefore assume they follow similar $T_e - T_e$ relations. We establish a correlation using the combined data set of CHAOS and point-sources, and then calculate the residuals of the diffuse gas sample relative to these $T_e - T_e$ relations to quantify any deviation.

\begin{figure*}
    \centering
    \includegraphics[width=0.8\textwidth]{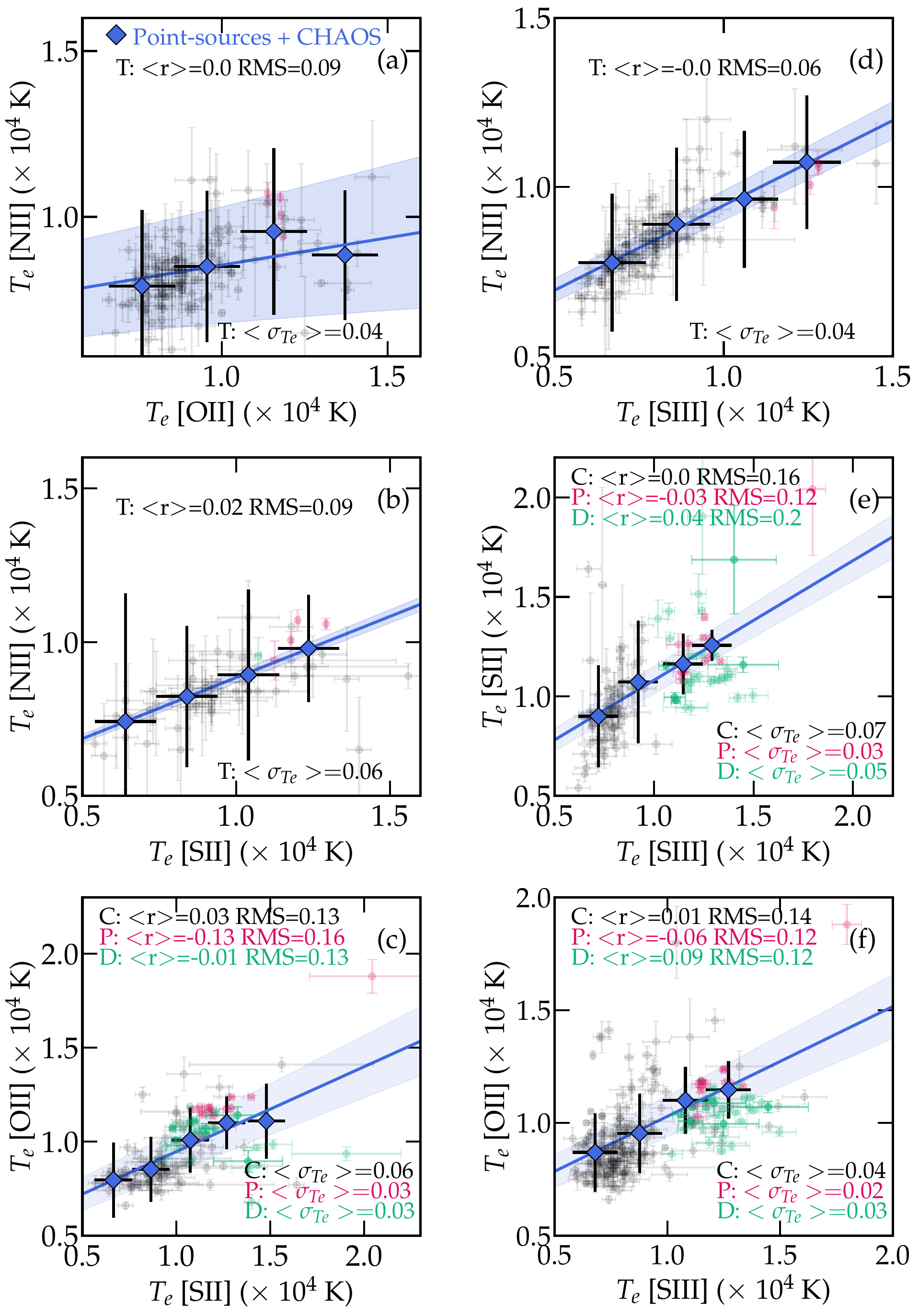}
    \caption{Same as Figure~\ref{fig:te-te_low_ion}, but with blue diamonds indicating bins that include CHAOS (black points) and point-like sources (pink points). The blue line shows the fit to the bins, and the blue shaded region shows the 1$\sigma$ uncertainty. Panels (a), (b), and (d) display the mean residual ($\langle r \rangle$), RMS, and mean measurement uncertainty ($\sigma_{T_e}$) for the total sample, while panels (c), (e), and (f) show these quantities for the three individual samples (C = CHAOS, P = point-sources, D = diffuse gas).}
    \label{fig:te-te_low_ion_bin}
\end{figure*}

\begin{figure*}
    \centering
    \includegraphics[width=0.8\textwidth]{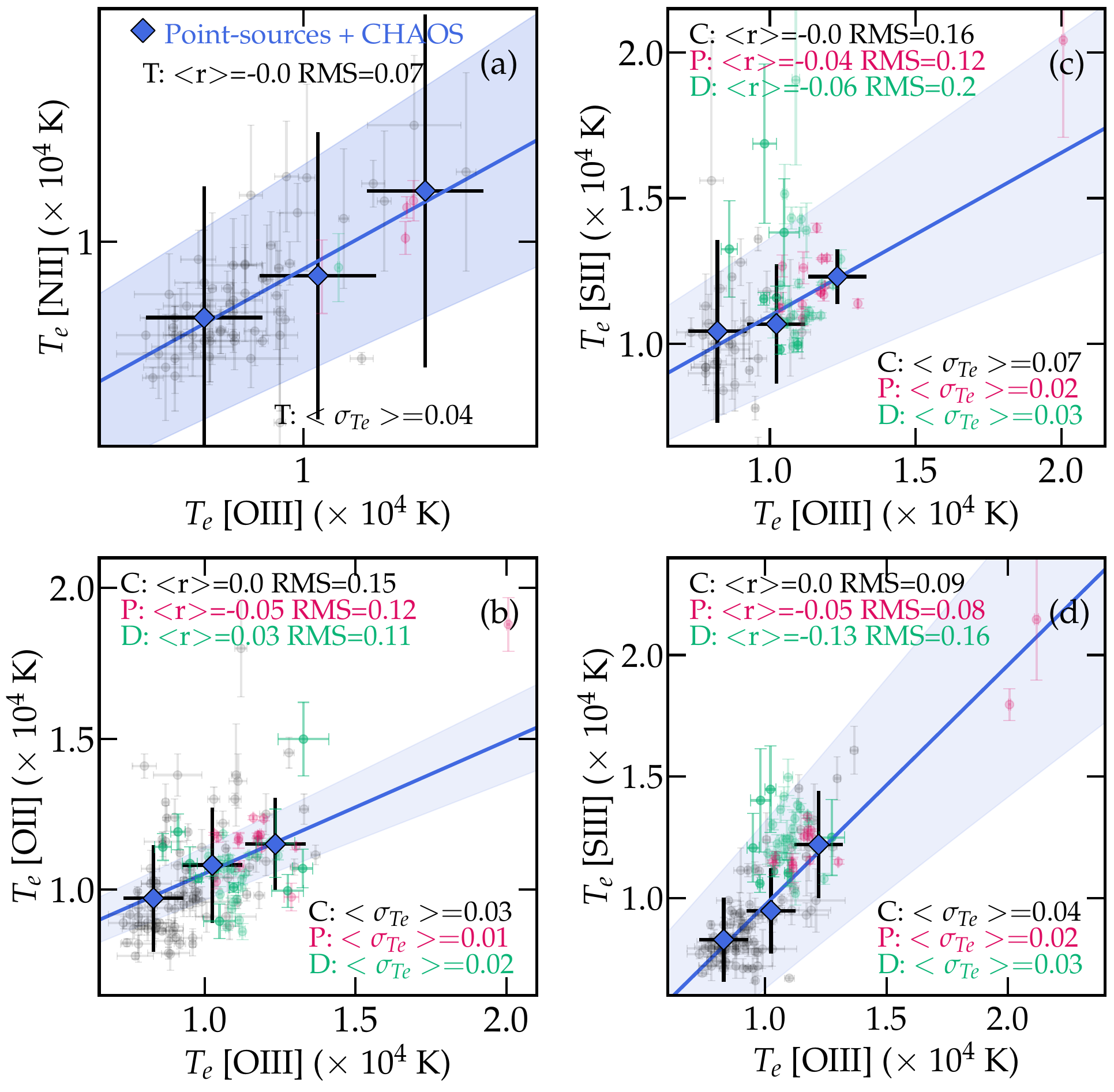}
    \caption{Same as Figure~\ref{fig:te-te_high_ion}, but with blue diamonds indicating bins that include CHAOS (black points) and point-like sources (pink points). The blue line shows the fit to the bins, and the blue shaded region shows the 1$\sigma$ uncertainty. Panel (a) displays the mean residual ($\langle r \rangle$), RMS, and mean measurement uncertainty ($\sigma_{T_e}$) for the total sample, while panels (b), (c), and (d) show these quantities for the three individual samples (C = CHAOS, P = point-sources, D = diffuse gas).}
    \label{fig:te-te_high_ion_bin}
\end{figure*}

Our sample is unevenly distributed, with many more data points in the CHAOS sample. Linear fits to data with uneven spacing can lead to fits that are dominated by positions where there is an over-density of measurements. We, therefore, bin the data along the x-axis and perform linear fits on the binned values. This approach reduces the influence of over-densities and ensures that regions with fewer measurements are properly represented in the fit. We define the bins to be 2000~K wide, starting from the measurement with the lowest $T_e$, and each bin contains at least two measurements. Bins with fewer measurements are excluded from the fit. For each bin we estimate the mean $T_e$ value along the y-axis. We also performed this test with the median value, and the results were unchanged. We adopt the width of the bin as the error bar in the x-axis and for the error bar in the y-axis we use the sum in quadrature of the standard deviation of the measurements and the mean uncertainty of the measurements.
Figures~\ref{fig:te-te_low_ion_bin}~\&~\ref{fig:te-te_high_ion_bin} show similar correlations as those in Figures~\ref{fig:te-te_low_ion}~\&~\ref{fig:te-te_high_ion} respectively. In these figures, we also plot the bins in blue. After binning, we fit a linear relationship to the bins using an orthogonal distance regression routine from \texttt{Python/SciPy}. In Figures~\ref{fig:te-te_low_ion_bin}~\&~\ref{fig:te-te_high_ion_bin} the blue shaded region represents the 1~$\sigma$ uncertainty of the fit to bins (illustrated with the blue line).  Black dots correspond to CHAOS, pink dots to point-like sources and green dots to diffuse gas measurements. We reiterate that bins only include CHAOS (black) and X-Shooter point-sources (pink). 

To estimate the residual of our measurements, we calculate the orthogonal distance from each point to the fitted line. We then calculate the mean residual, $\langle r \rangle$ and the RMS for each of the three samples around the fit. We do not have enough detections of [\ion{N}{II}] to carry out this analysis. We therefore only calculate the $\langle r \rangle$ and the RMS for all points combined for any relation that includes [\ion{N}{II}]. In Figures~\ref{fig:te-te_low_ion_bin} \& \ref{fig:te-te_high_ion_bin}, $\langle r \rangle$ and RMS are shown at the top of each panel, for each sample (T = total, C = CHAOS, P = point-like sources, D = diffuse gas). For comparison, at the bottom of each panel, we show the mean measured uncertainty ($\sigma_{Te}$) for each sample. In Section~\ref{sec:HII_vs_diffuse}, we discuss these values and interpret them in the context of comparing the different samples.

\label{append:binned te-te}

\bsp	
\label{lastpage}
\end{document}